%% file: paper.tex
\documentclass[journal]{IEEEtran}
\usepackage{url}
\usepackage{cite}
\usepackage{xspace}
\usepackage{multirow}
\usepackage{ifthen}
\usepackage{xcolor,colortbl}
\usepackage{tabulary}
\usepackage{balance}

\usepackage{chronosys}

\usepackage{tikz}
\usepackage{tikz-qtree}
\usetikzlibrary{trees}

\usepackage{textcomp}

\input{macros}

\usepackage{array}
\newcolumntype{L}[1]{>{\raggedright\let\newline\\\arraybackslash\hspace{0pt}}m{#1}}
\newcolumntype{C}[1]{>{\centering\let\newline\\\arraybackslash\hspace{0pt}}m{#1}}
\newcolumntype{R}[1]{>{\raggedleft\let\newline\\\arraybackslash\hspace{0pt}}m{#1}}

\makeatletter
\define@key{Gin}{Trim}
           {\let\Gin@viewport@code\Gin@trim\expandafter\Gread@parse@vp#1 \\}
           
\makeatother

\ifCLASSINFOpdf
\else
\fi

\usepackage[tight,footnotesize]{subfigure}

\begin{document}
\title{Low Power Wide Area Networks: An Overview}

\author{Usman Raza,
Parag Kulkarni, and
Mahesh Sooriyabandara
\thanks{All authors  are with  Toshiba Research Europe Limited, UK, Email: \{usman.raza, parag.kulkarni, mahesh.sooriyabandara\}@toshiba-trel.com.
\color{blue}{\copyright 2016  IEEE. Personal use  of this material is  permitted. Permission from  IEEE must be obtained  for all
other uses, in any current or future media, including reprinting/republishing this material for advertising or
promotional purposes, creating new collective works, for resale or redistribution to servers or lists, or reuse
of any copyrighted component of this work in other works.}
}
}

\maketitle


\input{abstract}


% \fi
\IEEEpeerreviewmaketitle

\input{intro}
\input{goals_techniques}

\section{Proprietary Technologies}
\label{sec:proprietary}

%Mapping the techniques described above with all the standards
%LPWA constantly evolving
%In this article, we identify several LPWA technologies, ranked in perceived importance, which will be crucial in future wireless standards.
In this section, we highlight and compare emerging proprietary technologies shown in Figure~\ref{fig:proprietary} and their technical aspects summarized in Table~\ref{tab:specifications}. Some of these technologies are being made compliant to the standards proposed by the different SDOs and SIGs. We dedicate Section~\ref{sec:standards} to briefly describe these standards and their association with any proprietary technologies discussed next. 

%Figure Taxonomy. 

%\tikzset{edge from parent/.style=
%{draw, edge from parent path={(\tikzparentnode.east)
%-- +(0,-8pt)
%-| (\tikzchildnode)}},
%blank/.style={draw=none}
%}

%
%\begin{figure}
%\begin{tikzpicture}[grow=right,level distance=1.25in,sibling distance=.25in]
%\matrix
%{
%\tikzset{edge from parent/.style= 
%            {thick, draw, edge from parent fork right},
%         every tree node/.style=
%            {draw,minimum width=1in,text width=1in,align=center}}
%
%\node{\Tree 
% [.{Proprietary Technologies for LPWA} 
%    [{\sigfox} {\lora} {\ingenu \rpma} {\qowisio} {\dash}  {\telensa} ]
% ]};\\
%};           
%\end{tikzpicture}
%\end{figure}

\begin{figure}
\begin{tikzpicture}[grow'=right,level distance=1.10in,sibling distance=.25in]
\tikzset{edge from parent/.style= 
            {thick, draw, edge from parent fork right},
         every tree node/.style=
            {draw,minimum width=0.5in,text width=0.8in,align=center}}
\Tree 
    [. {Proprietary LPWA Technologies}
        [{\sigfox} {\lora} {\ingenu \rpma} {\telensa} {\qowisio} ]
    ]
\end{tikzpicture}
\caption{Emerging proprietary LPWA technologies.}
\label{fig:proprietary}
\end{figure}

%\begin{figure*}
%\begin{tikzpicture}
%\matrix
%{
%\node{\Tree 
% [.{Low Power Wide Area Networks} 
%    [.{Proprietary Technologies} 
%    		[{\sigfox} {\lora} {\ingenu} {\qowisio} {\dash}  {\telensa} ]]
%    [.{Standards} 
%        [{IEEE 802.15.4k} {ETSI LTN} {eMTC} {\nbiot} {\lorawan} {\weightless(W,N,P)} ] ]
% ]};\\
%};           
%\end{tikzpicture}
%\end{figure*}

\input{specifications} % Table

\subsection{\sigfox}
\label{subsec:sigfox}
\sigfox itself or in partnership with other network operators offers an end-to-end LPWA connectivity solution based on its patented technologies. \sigfox Network Operators (SNOs) deploy the proprietary base stations equipped with cognitive software-defined radios and connect them to the backend servers using an IP-based network. The end devices connect to these base stations using Binary Phase Shift Keying (BPSK)  modulation in an ultra narrow (100Hz) \subghz \ism band carrier. By using UNB, \sigfox utilizes bandwidth efficiently and experiences very low noise levels, resulting in high receiver sensitivity\usman{-142dBm}, ultra-low power consumption, and inexpensive antenna design.  All these benefits come at an expense of maximum throughput of only 100 bps. The achieved data rate clearly falls at the lower end of the throughput offered by most other LPWA technologies and thus limits the number of use-cases for \sigfox. Further, \sigfox initially supported only uplink communication but later evolved into a bidirectional technology, although with a significant link asymmetry. The downlink communication can only precede uplink communication after which the end device should wait to listen for a response from the base station. The number and size of messages over the uplink are limited to 140 12-byte messages per day to conform to the regional regulations on use of license-free spectrum~\cite{spectrumuse}. Radio access link is asymmetric, allowing transmission of maximum of only 4 8-bytes per day over the downlink from the base stations to the end devices. It means that acknowledging every uplink message is not supported. 

% simple end-devices more complex base stations. no acks, no agreement on which channel to use. basestation listen on all channels. end devices do not listen acks. just multiple transmissions. 
Without adequate support for acknowledgments, reliability of the uplink communication is improved by using time and frequency diversity as well as redundant transmissions. A single message from an end device can be transmitted multiple times over different frequency channels. For this purpose, in Europe, the band between 868.180-868.220MHz is divided into 400 100Hz channels~\cite{waspmote}, out of which 40 channels are reserved and not used. As the base stations can scan all the channels to decode the messages, the end devices can autonomously choose a random frequency channel to transmit their messages. This simplifies the design for the end devices. Further, a single message is transmitted multiple times (3 by default) to increase the probability of successful reception by the base stations. 

\subsection{\lora}
\label{subsec:lora}
%For every technique
%Key features: modulation technique, operating band, data range range etc. and short description. 
%Place in taxonomy. 
%Status/development/deployment: Industry adoption.
\lora is a physical layer technology that modulates the signals in \subghz \ism band using a proprietary spread spectrum technique~\cite{lorapatent} developed and commercialized by Semtech Corporation~\cite{semtech}. A bidirectional communication is provided by a special chirp spread spectrum (CSS) technique, which spreads a narrow band input signal over a wider channel bandwidth. The resulting signal has noise like properties, making it harder to detect or jam. The processing gain enables resilience to interference~\cite{spreadinterferenceresilience} and noise. %of 125KHz\usman{that is not the only one!}.

%relaxes the accuracy needed for 

%still sufficient for a broad range of tracking and smart city applications. 
%to extends the range at expense of lower data rate. 
%Chirp spread spectrum modulation, although very popular in military domain, 
 
The transmitter makes the chirp signals vary their frequency over time without changing their phase between adjacent symbols. As long as this frequency change is slow enough so to put higher energy per chirp symbol, distant receivers can decode a severely attenuated signal several dBs below the noise floor\usman{number please}. \lora supports multiple spreading factors (between 7-12) to decide the tradeoff between range and data rate. Higher spreading factors delivers long range at an expense of lower data rates and vice versa. \lora also combines Forward Error Correction (FEC) with the spread spectrum technique to further increase the receiver sensitivity. The data rate ranges from 300 bps to 37.5 kbps depending on spreading factor and channel bandwidth. Further, multiple transmissions using different spreading factors can be received simultaneously by a \lora base station. In essence, multiple spreading factors provide a third degree of diversity after time and frequency. 

\memo {A few studies evaluated \lorawan in real world environments including outdoor~\cite{nolansigfoxvslora, lpwanloracoverage, oana} and even indoor~\cite{indoorfrance} settings. The work in~\cite{nolansigfoxvslora} evaluates \lora and \sigfox  through experiments carried out from a test deployment in Ireland. Findings indicate that a \lora base station deployed at 470 m above sea level could serve a coverage area of 1380 square kilometers in the test setup and that \sigfox technology was able to provide a 25 km test link between a client using 14 dBm and the base station with an signal to noise ratio consistently exceeding 20 dB being measured in the tests performed. Another study in~\cite{lpwanloracoverage} observed 15 km and 30 km communication ranges for \lorawan on ground  and water respectively in Oulu Finland. Furthermore, in another study ~\cite{ouluevaluation} conducted at a university, end devices transmitted at 14 dBm using highest spreading factor (12) to the base station that was located within 420 m radius. The packet delivery ratio  at the base station is recorded to be 96.7\%.  
}

The messages transmitted by the end devices are received by not a single but all the base stations in the range, giving rise to \emph{``star-of-stars''} topology. By exploiting reception diversity this way, \lora improves ratio of successfully received messages. However, achieving this requires multiple base stations in the neighborhood that may increase CAPEX and OPEX. The resulting duplicate receptions are filtered out in the backend system. Further, \lora exploits these multiple receptions of same message at different base stations for localization of the transmitting end device. For this purpose, a time difference of arrival (TDOA) based localization technique supported by very accurate time synchronization between multiple base station is used.
%time difference between   to localize the transmitting end device. %To the best of our knowledge, the localization accuracy for \lora is not yet reported from real 
%Time difference Absence of direct line-of-sight connection  low channel bandwidth and often

%As the coded signal reaches a distant receiver, it may attenuate its  power, often below the noise floor.  
%At very long distances, a \lora-based receiver is sensitive to decode the chirp signal correctly if it changes its frequency slowly enough. 

%The spreading factor can be increased to reach longer distances at an expense of lower data rate. 

%slower chirps contains more information per bit, easily decoded at long distances
%smaller bandwidth 

%A higher processing gain enabled by the spreading of the signal in frequency domain enables receiver to decode the received signal even below the noise floor.  

%This propriety modulation enables \lora to reach 15km and 2km in suburban and urban environments respectively.  

%Its latest specifications define rules governing use of regional bands for Europe, USA and China. \lora is designed specifically for battery-powered devices. It has been claimed that \lora enabled battery powered devices can achieve a 10-year lifetime if data is transmitted a few times per hour.  In LPWAN market, \lora is competing against \weightless, \sigfox, LTE-Cat M, IEEE 802.11ah, \dash and \nbiot etc.

A special interest group constituted by several commercial and industrial partners dubbed as \loraalliance proposed \lorawan, an open standard defining architecture and layers above the \lora physical layer. We briefly describe \lorawan under standards in Section~\ref{sec:standards}. 

\subsection{\ingenu \rpma}
\label{subsec:rpma}
\ingenu (formerly known as On-Ramp Wireless) proposed a proprietary LPWA technology, which unlike most other technologies does not rely on better propagation properties of \subghz band. Instead it operates in 2.4 GHz \ism band and leverages more relaxed regulations on the spectrum use across different regions~\cite{howrpmaworks, spectrumuse}. To offer an example, the regulations in USA and Europe do not impose a maximum limit on duty cycle for 2.4 GHz band, enabling higher throughput and more capacity than other technologies operating in \subghz band.  

Most importantly, \ingenu uses a patented physical access scheme named as Random Phase Multiple Access (\rpma)~\cite{rpmapatent} Direct Sequence Spread Spectrum, which it employs for uplink communication only. %\rpma is a variation of direct sequence spread spectrum (DSSS) modulation. 
%In contrast to conventional DSSS where multiple transmitters 
%more bandwidth 80MHz
%1MHz 
As a variation of Code Division Multiple Access (CDMA) itself, \rpma enables multiple transmitters to share a single time slot. However, \rpma first increases time slot duration of traditional CDMA and then scatters the channel access within this slot by adding a random offset delay for each transmitter. By not granting channel access to the transmitters exactly at once (i.e., at the beginning of a slot), \rpma reduces overlapping between transmitted signals and thus increases signal to interference ratio for each individual link~\cite{howrpmaworks}. On the receiving side, the base stations employ multiple demodulators to decode signals arriving at different times within a slot.  
%does not start multiple transmissions at exactly but add a random delay 
%to improves 
%Another notable difference between \rpma and CDMA is that the former uses exactly same sequence code for all the 
%Within a single time slot, RPMA desynchronizes the transmissions from multiple end devices by randomizing their start times. %   multiple end-devices introduces a random delay before accessing the medium in a given time slot.  
%On the receiver side, the received signal does not perfectly overlap with other interfering signals and thus signal to interference ratio (SINR) is improved, leading to a higher communication reliability and better scalability. \usman{you should explain this better!!!!} %and capacity (more than \lora and \sigfox).
%Due to this reason, \ingenu aids uplink more in terms of reliability compared to the downlink where code division multiple access (CDMA) is used by the base station to send data concurrently to multiple end-devices. The base station spreads the signals and then transmits them with maximal overlap in time, resulting in lower SINR due to a higher interference among the multiple signals transmitted concurrently. 
\ingenu provides bidirectional communication, although with a slight link asymmetry. For downlink communication, base stations spreads the signals for individual end devices and then broadcast them using CDMA. 

%The end-devices can report the messages to the base station at scheduled times as well as asynchronously. 

\rpma is reported to achieve up to -142 dBm receiver sensitivity and 168 dB link budget~\cite{howrpmaworks}. Further, the end devices can adjust their transmit power for reaching closest base station and limiting interference to nearby devices. 

%Use of random offsets in transmission time slots favors uplink communication more than downlink. 
%As a result, it is obvious that \ingenu introduces a slight link asymmetry where uplink communications are more reliable than the downlink communications. 

\ingenu leads efforts to standardize the physical layer specifications under IEEE 802.15.4k standard. \rpma technology is made compliant to the IEEE 802.15.4k specifications.  

%both scheduled updates
%asynchronous event messages

%Compared to CDMA where multiple coded signals perfectly overlap in time, the signal to interference ratio (SINR) is improved by RPMA~\cite{bristol}. 

%Like CDMA, transmissions from multiple end-devices overlap but are made not to start at the same time by introducing a random delay. 

\subsection{\telensa} 
\telensa~\cite{telensa} provides end-to-end solutions for LPWA applications incorporating fully designed vertical network stacks with a support for integration with third party software. 

For a wireless connectivity between their end devices and the base stations, \telensa designed a proprietary UNB modulation technique~\cite{telensapatent}, which operates in license-free \subghz \ism band at low data rates. While less is known about the implementation of their wireless technology, \telensa aims to standardize its technology using ETSI Low Throughput Networks (LTN) specifications for an easy integration within applications. %\telensa has integrated their UNB technology with a suite of M2M applications called \emph{aptos}\cite{aptos}. 
%pursuing standardization under ETSI for LPWA connectivity 

\telensa currently focuses on a few smart city applications such as intelligent lighting, smart parking, etc. To strengthen their LPWA offerings in intelligent lighting business,  \telensa is involved with TALQ consortium~\cite{talq} in defining standards for monitoring and controlling outdoor lighting systems.  

\subsection{\qowisio} 
\qowisio deploys dual-mode LPWA networks combining their own proprietary UNB technology with \lora. It provides LPWA connectivity as a service to the end users: It offers end devices, deploys network infrastructure, develops custom applications, and hosts them at a backend cloud. Less is however known about the technical specifications of their underlying UNB technology and other system components.

%*****************************************************************************
\section{Standards}
\label{sec:standards}

\begin{figure}
\includegraphics[scale=0.85]{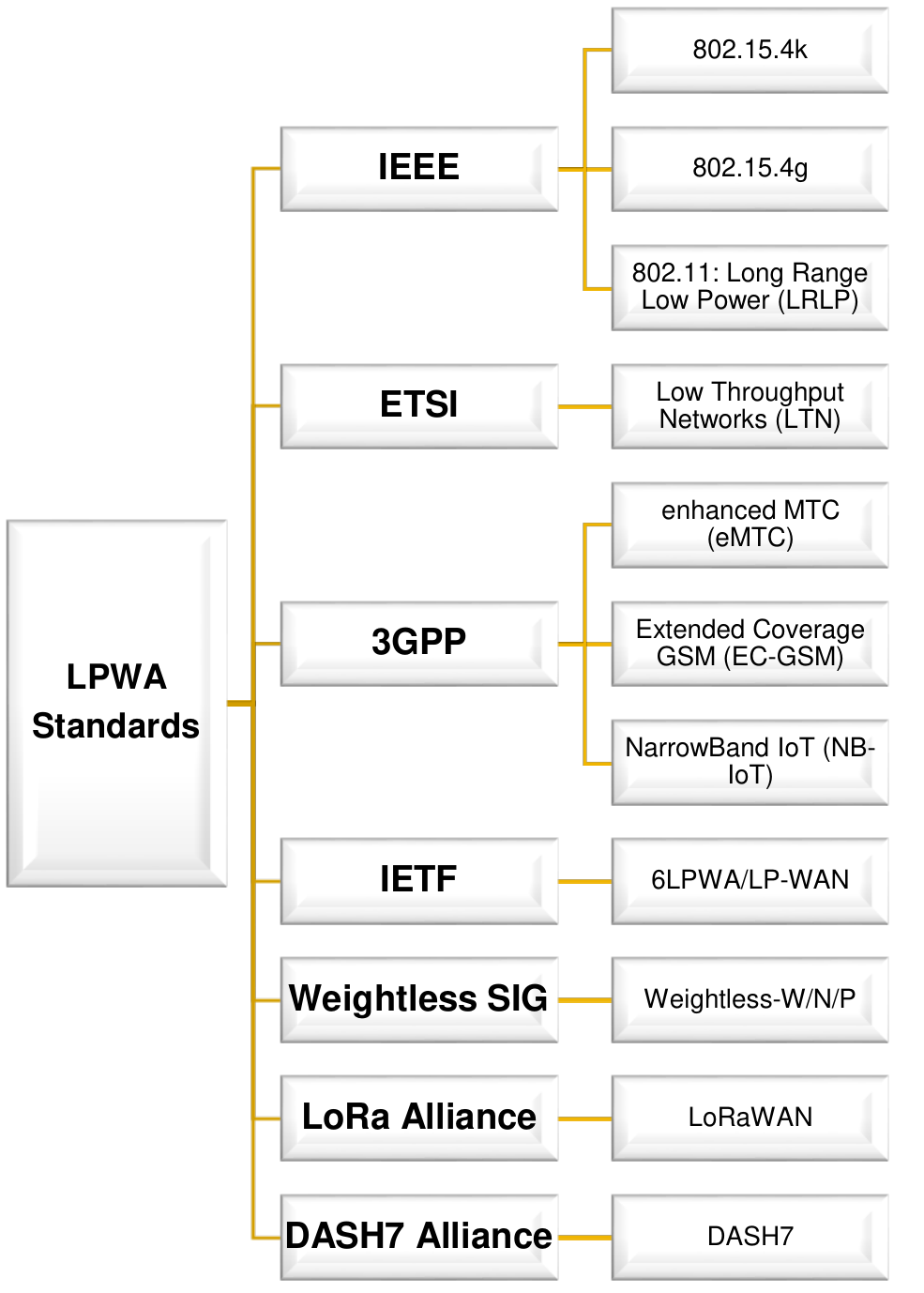}
\caption{LPWA standards and their developing organizations}
\label{fig:sdos}
\end{figure}

A plethora of standardization efforts are undertaken by different established standardization bodies including Institute of Electrical and Electronics Engineers (IEEE), European Telecommunications Standard Institute (ETSI), and The Third Generation Partnership Project (3GPP) along with  industrial consortia such as \weightless-SIG, \loraalliance, and \dash Alliance. Figure~\ref{fig:sdos} organizes the proposed standards according to their developing organizations, while Table~\ref{tab:standardspecifications} summarizes technical specifications of different standards. \memo{A qualitative comparison of some LPWA technologies can be found in~\cite{qualitativesurvey}}. Most of these efforts also involve several proprietary LPWA connectivity providers discussed in the previous section. The objectives of these SDOs and SIGs are quite diverse. In the long run, it is hoped that adoption of these standards will likely reduce the fragmentation of LPWA market and enable co-existence of multiple competing technologies. 
%. and ecosystem of co-existence and fair competition. 

\input{standardspecifications}

%\begin{figure*}
%%\begin{chronology}[5]{1983}{2010}{3ex}{4}
%%\event{1984}{one}
%%\event[1985]{1986}{two}
%%\event{\decimaldate{25}{12}{2001}}{three}
%%\end{chronology}
%\setupchronology{startyear=1000,color=blue,stopdate=false}
%%\setupchronoperiode{color=green}
%%\setupchronoperiode{color=\chronoperiodcolor}
%%\chronoperiodecoloralternation{orange, darkgreen, violet, purple, cyan}
%\chronoperiodecoloralternation{orange, green, violet, purple, cyan}
%\setupchronoevent{textstyle=\it}
%\setupchronograduation[event]{markdepth=2cm}
%\startchronology
%\chronograduation{250}
%\chronoperiode{1050}{1450}{Anything you want}
%\chronoevent{1600}{Anything else}
%\chronoperiode{1800}{1899}{$19^{th}$ century}
%\stopchronology
%\end{figure*}

\subsection{IEEE}

IEEE is extending range and reducing power consumption of their 802.15.4~\cite{802154} and 802.11~\cite{80211} standards with the set of new specifications for the physical and the MAC layers. Two LPWA standards are proposed as amendments to IEEE 802.15.4 base standard for Low-Rate Wireless Personal Area Networks (LR-WPANs), which we will cover in this section. Along with this, the efforts on amending IEEE  802.11 standard for wireless local area networks (WLANs) for longer range are also briefly described. 

\subsubsection{IEEE 802.15.4k: Low Energy, Critical Infrastructure Monitoring Networks.}
IEEE 802.15.4k Task Group (TG4k) proposes a standard for low-energy critical infrastructure monitoring (LECIM) applications to operate in the \ism bands (\subghz and 2.4 GHz). This was a response to the fact that the earlier standard falls short on range and the node densities required for LPWA applications. IEEE 802.15.4k amendment bridges this gap by  adopting DSSS and FSK as two new PHY layers. Multiple discrete channel bandwidths ranging from 100kHz to 1MHz can be used. The MAC layer specifications are also amended to address the new physical layers. The standard supports conventional CSMA/CA without priority channel access (PCA), CSMA, and \aloha with PCA. With PCA, the devices and base stations can prioritize their traffic in accessing the medium, providing a notion of quality of service. 
%The DSSS physical layer can vary its spreading factor between 16 and 32768 to provide interference resilient and long range communication. 
%Point to multi-thousands of point communication is supported 
Like most LPWA standards, end-devices are connected to the base stations in a star topology and are capable of exchanging asynchronous and scheduled messages.
%point to multi-thousands of point communication 
%priority channel access PCA

\memo {An IEEE 802.15.4k based LPWA deployment for air quality monitoring is elaborated in~\cite{lpwa11k}. A star topology network was deployed wherein 1 access point and 5 nodes were deployed within a 3 km radius area from the center of the university campus. The access point operates in the 433 MHz spectrum. Using a transmit power of 15 dBm the transceiver can support different sensitivities depending on the data rate requirements, e.g. sensitivities of -129 dBm, -123 dBm and -110 dBm can be achieved for data rates corresponding to 300 bps, 1.2 kbps and 50 kbps respectively. }

\ingenu, the provider of the \rpma LPWA technology~\cite{howrpmaworks}, is a proponent of this standard. The PHY and MAC layers of \ingenu's LPWA technology are compliant with this standard. 

\subsubsection{IEEE 802.15.4g: Low-Data-Rate, Wireless, Smart Metering Utility Networks}
IEEE 802.15 WPAN task group 4g (TG4g) proposes first set of PHY amendments to extend the short range portfolio of IEEE 802.15.4 base standard. The release of standard in April 2012~\cite{154g} addresses the process-control applications such as smart metering networks, which are inherently comprised of massive number of fixed end devices deployed across cities or countries. The standard defines three PHY layers namely FSK, Orthogonal Frequency-Division Multiple Access (OFDMA), and offset Quaternary Phase Shift Keying (QPSK), which support multiple data rates ranging from 40 kbps to 1 Mbps across different regions. With an exception of a single licensed band in USA, the PHY predominantly operates in \ism (\subghz and 2.4 GHz) bands and thus co-exists with other interfering technologies in the same band. The PHY is designed to deliver frames of size up to 1500 bytes so to avoid fragmenting Internet Protocol (IP) packets. 

The changes in the MAC layer to support the new PHYs are defined by IEEE 802.15.4e and not by IEEE 802.15.4g standard itself.

\subsubsection{IEEE 802.11: Wireless Local Area Networks}

\memo{WLAN technologies will play an important role in IoT~\cite{tozluwifi}}. The efforts for extending range and decreasing power consumption for WLANs are made by the IEEE 802.11 Task Group AH (TGah) and the IEEE 802.11 Topic Interest Group (TIG) in Long Range Low Power (LRLP). 

TGah~\cite{ah} proposed the IEEE 802.11ah specifications for PHY and MAC to operate long range Wi-Fi operation in \subghz \ism band. \memo{Compared to IEEE 802.11ac standard, several new features were introduced to achieve 1 km range in outdoor environments and the data rate in excess of 100 kbps. The PHY adopts OFDM that transmit at the rate 10 times slower than IEEE 802.11ac, an earlier standard, so to extend the communication range. At the MAC layer, overheads associated with frames, headers and beacons are reduced to prolong battery powered operation~\cite{anotherah}.  MAC protocol is tailored to thousands (8191) of connected end devices so that it reduces the resulting collisions among them. End devices are enabled with mechanisms to save energy during the inactive periods but yet  retain their connection/synchronization with the access points. With all these new power saving and range enhancements, IEEE 802.11ah indeed provides significantly longer range and lower energy consumption than other WLAN standards, ZigBee, and Bluetooth but not as much as the other LPWA technologies discussed in this paper. Due to this reason, increasing number of recently published studies~\cite{andreev2015understanding, landscape} and IETF draft documents~\cite{gapanalysis, ietflpwan} do not enlist IEEE 802.11ah as a LPWA technology. In fact, IEEE 802.11ah caters to those applications that require relatively higher bandwidth at the expense of higher power consumption than the other LPWA technologies. }

%While IEEE 802.11ah competes with LPWA technologies in IoT landscape~\cite{landscape}, it is not to be classified as such~\cite{landscape} due to its relatively shorter range and lack of support for mobility. 

%\memo{Low power WiFi caters to those applications that require relatively higher bandwidth at the expense of higher power consumption than other technologies discussed in this paper. 
%%Unlike other LPWA technologies like \sigfox, \lorawan and \ingenu, low power WiFi is not offered as a \emph{managed service} and therefore does not free customers from the hassle of managing their own WiFi infrastructure. 
%}

%offload cellular traffic, 
\memo{ Feasibility of using IEEE 802.11ah for IoT/M2M use cases is studied in~\cite{hazmi}. The authors show that when using the 900 MHz band, for the downlink case, it is straightforward to achieve a 1 km range and higher than 100 kbps data rate as the AP uses higher transmit power (20-30 dBm). However, for the uplink case, it is quite challenging to achieve these targets as the clients operate a low power (0 dBm) and are to be duty cycled to enable years of battery operation. In such a case, range of up to 400 m was achieved with the authors highlighting that use of coding schemes, higher transmit power and higher gain antennas could potentially help to improve this. However this may come at the cost of reduced battery life at the clients which may not be desirable. They also suggest that if the reliability requirements are reduced, range can be further increased, e.g. they were able to achieve 1 km range for a link reliability less than 60\%. } 

\memo{A new Topic Interest Group (TIG) was setup under the remit of 802.11 in 2016 to explore feasibility of a new standard for Long Range Low Power (LRLP)~\cite{lrlp}. At an early stage of this work, the TIG had defined some use cases and functional requirements for this technology in \cite{lrlpspecs} but could not clearly justify need for this activity within the IEEE LAN/MAN Standards Committee (LMSC). Therefore, the work on LRLP came to a premature end.}

\subsection{ETSI}
 ETSI leads efforts to standardize a bidirectional low data rate LPWA standard. The resulting standard dubbed as \emph{Low Throughput Network (LTN)} was released in 2014 in the form of three group specifications. These specifications define i) the use cases~\cite{ltn1} ii) the functional architecture~\cite{ltn2}, and iii) the protocols and interfaces~\cite{ltn3}. One of its primary objectives is to reduce the electromagnetic radiation by exploiting short payload sizes and low data rates of M2M/IoT communication. 

%These specifications lay the foundation of the standardization efforts coordinated by ETSI's Electromagnetic Compatibility and Radio Spectrum Matters committee (TC ERM). 
%intends to propose optimizations for low payload sizes (typically 12 byte) so to reduce electromagnetic radiation. 

Apart from the recommendation on the air interfaces, LTN defines various interfaces and protocols for the cooperation between end-devices, base stations, network server, and operational and business management systems. 

Motivated by the fact that the emerging LPWA networks use both ultra narrow band (e.g.,\sigfox, \telensa) and orthogonal sequence spread spectrum (OSSS) (e.g., \lora) modulation techniques, LTN standard does not restrict itself to a single category. It provides flexibility to LPWA operators to design and deploy their own proprietary UNB or OSSS modulation schemes in \subghz \ism band as long as the end-devices, base stations and the network servers implement the interfaces described by the LTN specifications~\cite{ltn1, ltn2, ltn3}. These specifications recommend using BPSK in uplink and GFSK in downlink for a UNB implementation. Alternatively, any OSSS modulation scheme can be used to support bidirectional communication. Data encryption as well as user authentication procedures are defined as a part of the LTN specifications.

Several providers of LPWA technologies such as \sigfox, \telensa, and Semtech are actively involved with ETSI for standardization of their technologies. 

%allows interfaces based on OSSS or UNB in license exempt sub-Ghz band. While implementing UNB, recommnds using BPSK in uplink  and GFSK in the downlink. While UNB is used for most uplink communications the end-device can choose to received downlink communication by agreeing on a window with basestation. 
%
% 
% 
% keep options open for differenet LPWA technologies to either implement UNB or OSSS modulation techniuqe. In essence, it is superset of features radio interface features available present in \sigfox and \lora.
% Yet as a standard, it allows other LPWA contenders to  propose their own  
% superset of features 

\subsection{3GPP}
\label{sec:3gpp}

%massive IoT vs critical IoT applications. 
%3GPP is drieving standardization in to diversify its cabailities
%3GPP is stripping complexity off from the cellular standards to make it more suitable for machine type communication. 
To address M2M and IoT market, 3GPP is evolving its existing cellular standards to strip complexity and cost, improve the range and signal penetration, and prolong the battery lifetime. Its multiple licensed solutions such as  Long Term Evolution (LTE) enhancements for Machine Type Communications (eMTC), Extended Coverage GSM (EC-GSM), and Narrow-Band IoT (\nbiot) offer different trade-offs between cost, coverage, data rate, and power consumption to address diverse needs of IoT and M2M applications. However, a common goal of all these standards is to maximize the re-use of the existing cellular infrastructure and owned radio spectrum. 
%integrate easily current cellular infrastructure and evolve in the road towards 5G and emergence of new use cases. 

\subsubsection{LTE enhancements for Machine Type Communications (eMTC)}
Conventional LTE end devices offer high data rate services at a cost and power consumption not acceptable for several MTC use cases. To reduce the cost while being compliant to LTE system requirements, 3GPP reduces the peak data rate from LTE Category 1 to  LTE Category 0 and then to LTE Category M, the different stages in the LTE evolution process. Further cost reduction is achieved by supporting optional half duplex operation in Category 0. This choice  reduces the complexity of modem and antenna design. From Category 0 to Category M1 (also known as eMTC), a more pronounced drop in the receive bandwidth from 20 MHz to 1.4 MHz in combination with a reduced transmission power will result in more cost-efficient and low-power design. %The Cat-M is due in release 
%along with a support for discontinued 

% energy saving mechanisms: extended Discontinous Reception (eDRX): end device do not monitor the down link control channel during some time period.  

To extend the battery lifetime for eMTC, 3GPP adopts two features namely \emph{Power Saving Mode (PSM)} and \emph{extended Discontinuous Reception (eDRx)}. They enable end devices to enter in a deep sleep mode for hours or even days without losing their network registration. The end devices avoid monitoring  downlink control channel for prolonged periods of time to save energy. The same power saving features are exploited in EC-GSM described next. 
 
\subsubsection{EC-GSM}
While Global System for Mobile Communications (GSM)  is announced to be decommissioned in certain regions, Mobile Network Operators (MNOs) may like to prolong their operation in few markets. With this assumption, 3GPP is in process of proposing the extended coverage GSM (EC-GSM) standard that aims to extend the GSM coverage by +20dB using \subghz band for better signal penetration in indoor environments. A link budget in the range of 154 dB-164 dB is aimed depending on the transmission power\usman{23-33dBm}. With only a software upgrade of GSM networks, the legacy GPRS spectrum can pack the new logical channels defined to accommodate EC-GSM devices. EC-GSM exploits repetitive transmissions and signal processing techniques to improve coverage and capacity of legacy GPRS.  Two modulation techniques namely Gaussian Minimum Shift Keying (GMSK) and 8-ary Phase Shift Keying (8PSK) provide variable data rates with the peak rate of 240 kbps with the latter technique. The standard was released in mid 2016 and aims to support 50k devices per base station and enhanced security and privacy features compared to conventional GSM based solutions. 

%http://www.3gpp.org/images/presentations/3GPP_Standards_for_IoT.pdf 
%decommissioning to reclaim the spectrum for more efficient 4G and 5G. However, where gsm will stay software based upgrade.
\subsubsection{\nbiot}

\memo{
\nbiot is a narrow-band technology that was made available as a part of Release-13 around mid 2016. \nbiot aims at enabling deployment flexibility, long battery life, low device cost and complexity and signal coverage extension. \nbiot is not compatible with 3G but can coexist with GSM, GPRS and LTE. \nbiot can be supported with only a software upgrade on top of existing LTE infrastructure. It can be deployed inside a single GSM carrier of 200 kHz, inside a single LTE physical resource block (PRB) of 180 kHz or inside an LTE guard band. Compared to eMTC, \nbiot cuts the cost and energy consumption further by reducing the data rate and bandwidth requirements (needs only 180 kHz) and simplifying the protocol design and mobility support. Further, a standalone deployment in a dedicated licensed spectrum is supported. 

\nbiot aims for a 164 dB coverage, serving up to 50k end devices per cell with the potential for scaling up the capacity by adding more \nbiot carriers. \nbiot uses single-carrier Frequency Division Multiple Access (FDMA) in uplink and Orthogonal FDMA (OFDMA) in downlink~\cite{dinoflore16}. The data rate is limited to 250 kbps for the multi-tone downlink communication and to 20 kbps for the single-tone uplink communication. As highlighted in \cite{APrimerOn3GPPNarrowBandIoT}, for a 164 dB coupling loss, an \nbiot based radio can achieve a battery life of 10 years when transmitting 200 bytes of data per day on average. For an in-depth look into \nbiot, we refer the interested reader to \cite{APrimerOn3GPPNarrowBandIoT}. Further \cite{dinoflore16} compares the different cellular based LPWA options covered in this section. 

Further to publication of Release-13 specifications, \nbiot standard has been critiqued in \cite{nbiot7surprises}). We summarize this critique as follows:
\begin{itemize}
\item	Only half the messages are acknowledged in NB-IoT due to limited downlink capacity. This implies the inability to realize IoT applications that require acknowledging of all uplink data traffic unless the application implements some form of reliability mechanisms. The latter could result in increased application complexity and higher energy consumption due to extra processing.
\item Use of packet aggregation (combining multiple packets and sending them as a single larger packet) in 3GPP based solutions improves efficiency but comes at the cost of extra latency that may be undesirable for delay sensitive IoT applications.
\item \nbiot traffic is best effort and therefore during times of heavy voice/data traffic, dynamically reallocating spectrum to relieve congestion for the latter class of traffic may impact \nbiot application performance. Further, once deployed an \nbiot device is likely to stay put for 10-20 years, an order of magnitude higher device upgrade cycle when compared to traditional mobile phones (typically 2 years). Some applications may take longer to break-even and provide a return on investment. Moreover, if new cellular generations come along, there could be questions with respect to the longevity of the deployed solution, e.g. a situation similar to some operators phasing out their GSM networks to reclaim the spectrum for LTE. This could leave the customers stranded since it may not be trivial/economically feasible to upgrade the end points, a valid argument. 
\item The lack of commercial deployments leaves open questions on the actual battery life and performance attainable in real world conditions.
%\item Whilst 3GPP estimates point to a 200-2000 bytes for a firmware upgrade, Ingenu is counter-arguing that these figures are not realistic in light of their deployment experiences. However, as there is no data to confirm/invalidate these claims given the lack of commercial deployments, time will tell. 
\end{itemize}
}

%\nbiot is a narrow-band technology that was initially proposed in September 2015 and is currently undergoing an active standardization process. \nbiot can be supported with only a software upgrade on top of existing LTE infrastructure. Compared to eMTC, \nbiot cuts the cost and energy consumption further by reducing the data rate and bandwidth requirements and simplifying the protocol design and mobility support. It requires a bandwidth of 180 KHz that can either come from in-band deployment with LTE or the unused guard bands. Further, a standalone deployment in a dedicated licensed spectrum is supported. 
%
%\nbiot aims for a 164 dB coverage, serving up to 50k end devices per cell. \nbiot uses single-carrier Frequency Division Multiple Access (FDMA) in uplink and Orthogonal FDMA (OFDMA) in downlink~\cite{dinoflore16}. The data rate is limited to 250kbps for the multi-tone downlink communication and to 20 kbps for the single-tone uplink communication. 
%% both support a data rate of up to 250kbps
%The complete PHY and MAC specifications of \nbiot are expected to be released in June 2016 as a part of Release 13.

\usman{By the end of 2016, complete large scale trail tests, More commercial activity will start thereon }

%reduce the cost of devices 10 times compared to 3GPP category 4
%stripped downed the hardware  
%will be achieved by reducing bandwidth, TX power, peak rate, memory requirement

%Sub-GHz bands have been preferred for deployments. discuss upgrade paths. For those 
%LTE 900 MHz
%LTE 800 MHz
%GSM 900 MHz upgrade.    

\subsection{IETF}

%The LPWA technologies surveyed in Section\ref{sec:proprietary} use properitary physical and link layer technologies, which are not compliant with each other. 
%with each other. 
%stresses the need to connect using standards. 
%The Internet 
%LPWA are not compliant with each other 
IETF aims to support LPWA ecosystem of dominantly proprietary technologies by standardizing \emph{end-to-end IP-based connectivity} for ultra-low power devices and applications. %Its main objective is to manage and secure end-devices, base stations, and network servers in a scalable fashion using mechanisms that can operate within very stringent constraints of underlying LPWA technologies. 
%Building a low-power network stack is 
IETF has already designed the IPv6 stack for Low power Wireless Personal Area Networks (6LoWPAN). However, these standardization efforts focus on legacy IEEE 802.15.4 based wireless networks, which support relatively higher data rates, longer payload sizes and shorter ranges than most LPWA technologies today.  %in terms of data rates, MTU, and time on air.  
%standardization to connect LPWA devices with outside word  well as others , while guaranteeing secure communication and
%connect each device 
%secure communication 
%manage devices
However, distinct features of LPWA technologies pose real technical challenges for the IP connectivity. Firstly, LPWA technologies are heterogeneous: every technology manipulates data in different formats using different physical and MAC layers. Secondly, most technologies use the \ism bands, which are subject to strict regional regulations, limiting maximum data rate, time-on-air, and frequency of data transmissions. Third, many technologies are characterized by a strong link asymmetry between uplink and downlink, usually limiting downlink capabilities. Thus, the proposed IP stacks should be lightweight enough to confine within these very strict limitations of the underlying technologies. Unfortunately, these challenges are not yet addressed in earlier IETF standardization efforts. 

%cope up with all these extra challenges, which are never addressed in earlier IETF standards. 
%All these extra challenges are to be coped with in providing IP connectivity to LPWA devices. 
%addressing, 
%Each LPWA technology differ from other 
%IETF has recently started looking into standardization of a network stack to connect the LPWA end devices to the Internet. While the need for such open standards cannot be stressed enough, heterogenity and the very constrianed nature of LPWA devices poses several technical challenges. 
%The 
%co-existence
A working group on Low-Power Wide Area Networks (LPWAN)~\cite{ietflpwan} under IETF umbrella was formed in April 2016. This group identified challenges and the design space for IPv6 connectivity for LPWA technologies in~\cite{gapanalysis}. Future efforts may likely culminate into multiple standards defining a full IPv6 stack for LPWA (6LPWA) that can connect LPWA devices with each other and their external ecosystem in a secure and a scalable manner. More specific technical problems to be addressed by this IETF group are described as follows:

%Efforts span different working groups is application signaling, MAC level scheduling,  
%
%6lo 6TiSCH, CoAP \usman{add a diagram of current standardization efforts.}
%but LPWA radically more stringent than 
%challenges: only a few short length messages can be exchanged,
%most techniques do not provide fragmentation support at layer 2
%lean protocol design for LPWA small code size 
%
%
%To connect low power devices to the Internet, IETF's existing standardization efforts focus on IEEE 802.15.4 based wireless personal area networks. Specifically, 
%
%has not addressed such low data rates and packet sizes. To fit an IP stack,
%such as 6LoWPAN 

%IETF LR-WAN group connecting heterogenous LPWA technology using Internet Protocol (IP) in  a secure and scalable manner. addressing, within stringent budgets.
%gaols: adaption layer that allows transportation of IPv6 packet over LPWA.

\begin{itemize}
\item \fakeparagraph{Header compression.} The maximum payload size for LPWA technologies is limited. The header compression techniques should be tailored to these small payload sizes as well as sparse and infrequent traffic of LPWA devices. %  Further for The maximum transmission unit varies with modulation rate .
\item \fakeparagraph{Fragmentation and reassembly.} Most LPWA technologies do not natively support fragmentation and reassembly at Layer 2 (L2). Because IPv6 packets are often too big to fit in a single L2 packet, the mechanisms for fragmentation and reassembly of IPv6 packets are to be defined. 
\item \fakeparagraph{Management.} To manage end devices, applications, base stations, and servers, there is a need for ultra-lightweight signaling protocols, which can operate efficiently over the constrained L2 technology. To this effect, IETF may look into efficient application-level signaling protocols~\cite{cosol}.  
%\item \fakeparagraph{multi-channel scheduling}: 
\item \fakeparagraph{Security, integrity, and privacy.} The IP connectivity should preserve security, integrity, and privacy of data exchanged over LPWA radio access networks and beyond.  Most LPWA technologies use symmetric key cryptography, in which end devices and the networks share the same secret key. More robust and resilient techniques and mechanisms may be investigated. 
\end{itemize}

\subsection{\loraalliance}
As described in Section~\ref{sec:proprietary}, \lora is a proprietary physical layer for LPWA connectivity. However, the upper layers and the system architecture are defined by \loraalliance under \lorawantm Specification~\cite{lorawan} that were released to public in July 2015. 

A simple \aloha scheme is used at the MAC layer that in combination with \lora physical layer enables multiple devices to communicate at the same time but using different channels and/or orthogonal codes (i.e., spreading factors). End devices can hop on to any base station without extra signaling overhead. The base stations connect end devices via a backhaul to network server, the brain of the \lorawan system that suppresses duplicate receptions, adapts radio access links, and forwards data to suitable application servers. Application servers then process the received data and perform user defined tasks. 

%replicating some of the features of features of IEEE 802.15.4 MAC to leaverage 
%As mobile end-devices communicate with all the gateways in the range, no handover is required.   

\lorawan anticipates that the devices will have different capabilities as per application requirements. Therefore, \lorawan defines three different classes of end-devices, all of which support bidirectional communication but with different downlink latency and power requirements. Class A device achieves the longest lifetime but with the highest latency.% for battery-powered sensors
It listens for a downlink communication \emph{only shortly after} its uplink transmission. 
%For this class, all the downlink communications should always follow an uplink transmission from the end device. 
%In addition, Class B device can even receive scheduled downlink communication to .
Class B device, in addition, can \emph{schedule} downlink receptions from base station at certain time intervals. Thus, only at these agreed-on epochs, applications can send control messages to the end devices (for possibly performing an actuation function). Lastly, Class C device is typically mains-powered, having capability to \emph{continuously} listen and receive  downlink transmissions with the shortest possible latency at any time. 

\lorawan standard uses symmetric-key cryptography\usman{AES} to authenticate end devices with the network and preserve the privacy of application data. 

\usman{
155 dB link budget.
125-500 kHz bandwidth
250bps - 50kbps in EU 
980bps-21.9kbps
multi-channel multi-modem basestations
}

%Each end-device can use a 128-bit AES encryption key to encrypt communication between itself and network server, while another 128-bit AES encryption can provide encryption between end-device and the application server. This comes handy if then end-users want to hide the information from the LoRaWAN service provider.  

\subsection{\weightless-SIG}
\weightless Special Interest Group~\cite{weightless} proposed three open LPWA standards, each providing different features, range and power consumption. These standards can operate in license-free as well as in licensed spectrum. 

\fakeparagraph{\weightless-W} leverages excellent signal propagation properties of TV white-spaces. It supports several modulation schemes including 16-Quadrature Amplitude Modulation (16-QAM) and Differential-BPSK (DBPSK) and a wide range of spreading factors. Depending on the link budget, the packets having sizes in upwards of 10 bytes can be transmitted at a rate between 1 kbps and 10 Mbps. The end devices transmit to base stations in a narrow band but at a lower power level than the base stations to save energy. \weightless-W has a one drawback. The shared access of the TV white spaces is permitted only in few regions, therefore \weightless-SIG defines the other two standards in \ism band, which is globally available for shared access. 

\fakeparagraph{\weightless-N} is a UNB standard for only one-way communication from end devices to a base station, achieving significant energy efficiency and lower cost than the other \weightless standards.  It uses DBPSK modulation scheme in \subghz bands. One-way communication, however, limits the number of use cases for \weightless-N.

\fakeparagraph{\weightless-P} blends two-way connectivity with two non-proprietary physical layers. It modulates the signals using GMSK and Quadrature Phase Shift Keying (QPSK), two well known schemes adopted in different commercial products. Therefore, the end devices do not require a proprietary chipset. Each single 12.5 kHz narrow channel in \subghz \ism band offers a data rate in the range between 0.2 kbps to 100 kbps. A full support for acknowledgments and bidirectional communication capabilities enable over-the-air upgrades of firmware. 

Like \lorawan, all \weightless standards employ symmetric key cryptography for authentication of end devices and integrity of application data. 

\subsection{\dash Alliance} 
The \dash Alliance is an industry consortium that defines a full vertical network stack for  LPWA connectivity known as \dash Alliance Protocol (D7AP)~\cite{dash7}. With its origin in the ISO/IEC 18000-7 standard~\cite{isoiec2009} for the air interface for active radio frequency identification (RFID) devices, D7AP has evolved into a stack that provides \emph{mid-range} connectivity to low-power sensors and actuators~\cite{dash7}. 

\dash employs narrow band modulation scheme using two-level GFSK in \subghz bands. Compared to most other LPWA technologies, \dash has a few notable differences. First it uses a  tree topology by default with an option to choose star layout as well. In the former case, the end devices are first connected to duty-cycling \emph{sub-controllers}, which then connect to the always ON base stations. This duty cycling mechanism brings more complexity to the design of the upper layers. %Second, \dash divides available spectrum into the channels of multiple distinct spacings to offer different data rates in the range between 9.6 kbps and 166.7 kbps .  
Second, \dash MAC protocol forces the end devices to check the channel periodically for  possible downlink transmissions, adding significant idle listening cost. By doing so, \dash gets much lower latency for downlink communication than other LPWA technologies but at an expense of higher energy consumption. Third, unlike other LPWA technologies, \dash defines a complete network stack, enabling applications and end devices to communicate with each other without having to deal with intricacies of the underlying physical or MAC layers. 
%low latency downlink communication to 

\dash implements support for forward error correction and symmetric key cryptography. 
%An open source implementation of full \dash stack is freely available. 
%maximum payload size 256 bytes

%command-response: downlink communication comes after uplink communication. 

%ISO 18000-7 -- \dash mode 2 -- > \dash alliance mode D7A

%An open source stack implementation of D7AP is freely available. 

%*****************************************************************************
\section{Challenges and Open Research Directions}
\label{sec:challenges}

\usman{Revolution, It is important to recognize the challenges at its onset, so to concentrate our efforts) }

%The work on LPWA networks is in its infancy. 
LPWA players are striving hard to innovate solutions that can deliver the so-called \emph{carrier grade performance}. To this effect, device manufacturers, network operators, and system integration experts have concentrated their efforts on cheap hardware design, reliable connectivity, and full end-to-end application integration. On the business side, the proprietary solution providers are in a rush to bring their services to the market and capture their share across multiple verticals. In this race, it is easy but counter-productive to overlook important challenges faced by LPWA technologies. In this section, we highlight these challenges and some research directions to overcome them and improve performance in long-term. 

\subsection{Scaling networks to massive number of devices}
%Techniques to scale LoRaWAN for a large number of end-devices and gateways  

LPWA technologies will connect tens of millions of devices transmitting data at an unprecedented scale over limited and often shared radio resources. This complex resource allocation problem is further complicated by several other factors. First, the device density may vary significantly across different geographical areas, creating the so called \emph{hot-spot} problem. These hot-spots will put the LPWA base stations to a stress test. Second, cross-technology interference can severely degrade the performance of LPWA technologies. This problem is definitely more severe for LPWA technologies operating in the license-exempt and shared \ism bands. Even licensed cellular LPWA technologies operating in-band with broadband services (like voice and video) are equally at this risk. It is not difficult to imagine a scenario when multiple UNB channels of a  LPWA technology are simultaneously interfered by a single broadband signal. Further, most LPWA technologies use simple \aloha or CSMA based MAC protocols, which do not scale well with number of connected devices~\cite{goodbyealoha}. 

%A similar problem can affect the cellular LPWA operating in-band with other legacy cellular services like voice, video, and data services on the licensed spectrum. 
%It is not yet understood how the LPWA technologies will solve complex resource allocation problems for huge number of devices. 
%Multiple non-LPWA technologies sharing unlicensed ISM adds another dimension 
%already a problem with technologies operating in shared ISM bands. However, in our opinion, cellular technologies will be no exception. 
%Existing LPWA technologies are not yet evaluated at such a large scale to  accurately quantify the scalability of these networks. Several research directions
%Massive number of IoT devices, each accounting for small traffic load, transmit a large cumulative traffic over shared radio resources. 

%Scalable to higher data rates per node, user densities, 

%Research directions:
%----------------------------

\memo{Multiple recent studies~\cite{lancasterlorascale, trllorascale, capacityscalability} investigate if LPWA technologies will be able to support large number of end-devices expected in future city-scale and nationwide deployments. At the time of writing only a few studies are present for \lorawan. Bor et al.~\cite{lancasterlorascale} estimate limit on number of nodes that can be supported by a typical \lorawan deployment to be 120 per 3.8 ha, a device density far less than expected in urban environments. Georgiou \& Raza~\cite{trllorascale} further unveil that \lorawan's coverage probability decay exponentially with number of end devices due to interference. Both studies seem to suggest that end devices should adapt \lora communication parameters possibly with help from more powerful base stations and exploit base station diversity to overcome this limitation.}

Several research directions can be pursued to address the capacity issue of LPWA technologies. These include use of channel diversity, opportunistic spectrum access, and adaptive transmission strategies. Use of channel hopping and multi-modem base stations can exploit channel and hardware diversity and is considered already for existing LPWA technologies. Cross-layer solutions can adapt the transmission strategies to the peculiar traffic patterns of LPWA devices and mitigate the effect of cross-technology interference. Further, improvements in existing MAC protocols are required for LPWA technologies to scale them well  for a large number of devices transmitting only short messages~\cite{goodbyealoha}. 

In the context of cellular LPWA networks, if excessive IoT/M2M traffic starves the legacy cellular traffic, MNOs may consider deploying LPWA support in unlicensed spectrum. Such an opportunistic use of radio spectrum can benefit from use of cognitive software-defined radios (SDR). SDRs could come in handy when multiple technologies need to compete for shared spectrum.

To cater to areas with a higher device density, LPWA access networks can borrow densification techniques from cellular domain. \memo {However, peculiarities of LPWA technologies such as their specialized modulation techniques, strong link asymmetry and mostly uncoordinated operation of end devices pose serious challenges to keep interference levels low in dense deployments.}

%The interference between several heterogeneous cell types require novel solutions. 

\subsection{Interference Control and Mitigation}
In future, the number of connected devices will undergo exponential increase, causing higher levels of interference to each other. The devices operating in the shared \ism bands will undergo unprecedented levels of both cross-technology interference as well as self-interference. Some interference measurement studies~\cite{interferencemeasurements} already point to a possible negative effect on coverage and capacity of LPWA networks. Furthermore, many LPWA technologies like \lora and \sigfox resort to simple \aloha scheme to grant channel access to the low-power end-devices. This choice of talking randomly without listening to others cannot only deteriorate performance, but also generates higher interference~\cite{randomnesscausingstrangeinterference}. Further, densification of the base station deployments to accommodate more devices is a major source of interference across LPWA cells and requires careful deployment  and design of base stations~\cite{utz}.

%with highly sensitive receivers of LPWA, the perceived interference will be non-negligible. 
In an anarchy of tens of wireless technologies and massive number of devices, all sharing the same channels, interference resilient communication and efficient spectrum sharing~\cite{policies} are key problems, both at technical and regulatory grounds. As interference varies across frequency, time, and space, devices should adapt their transmission schedules to experience the least interference and the best reliability. PHY and MAC layer designs exploiting this diversity at such a large scale need further investigation. Regulatory authorities may also need to step forward to propose rules to enable efficient sharing and cooperation between different wireless technologies in the unlicensed bands~\cite{policies}.
 
%available channels in sub-GHz band are limited , h

\subsection{High data-rate modulation techniques}
The LPWA technologies compromise on data rates to reach long distances. Some technologies especially those using UNB modulation in the shared \ism bands offer very low data rates and short payload sizes, limiting their potential business use cases. To support bandwidth hungry use cases, %requiring %higher data rates at least in certain application contexts, 
 it is meaningful to implement multiple modulation schemes for devices. As per application needs, devices can switch between different modulation schemes so to enable high energy efficiency, long range and high data rate simultaneously. 

To achieve this, there is a need for flexible and inexpensive hardware design that can support multiple physical layers, each of which can offer complementary trade-offs to match the range and data rate requirements of applications. 

%It suggest that evolving 3GPP efforts may have dual mode of operation. Or  multiple level of functionalities by weightless etc. 
%Weightless multiple standards
%Asymmetric latency as well

\subsection{Interoperability between different LPWA technologies}

\usman{It may make sense to split this into two: 1) connecting them to the Internet ---- minimize number and size of message exchanges, secure and radio access technology independent communication, }

%Heterogeneity of these technologies combined with an exponential increase in number of IoT devices ask for a strong interoperability support, which clearly lacks in current proprietary solutions. 
Given that market is heading towards an intense competition between different LPWA technologies, it is safe to assume that several may coexist in future. Interoperability between these heterogeneous technologies is thus crucial to their long-term profitability. With little to no support for interoperability between different technologies, a need for standards that glue them together is strong. Some of the standardization efforts across ETSI, IEEE, 3GPP, and IETF discussed in Section~\ref{sec:standards} will look into these interoperability issues. 
%virtualization, cloud computing, access technology independent IoT platforms, billing management  are likely the ways to ahead 
%To this effect, they are already pushing standards through the standardization bodies. 

%For interoperability, several research and design choices can be explored.%, each having its own implications on the network stacks running on end-devices, access-network, and backhaul.  
However, for a complete interoperability, several directions should be explored. Firstly, IP can already connect short-range wireless devices using mesh networking. The peculiarities of LPWA technologies limit a direct implementation of the same IP stack on LPWA devices. Alternative solutions based on gateways or backend based solutions are viable candidates. However, all such solutions should scale well with number of devices without degrading performance. Secondly, use of IoT middleware and virtualization techniques can play a major role in connecting LPWA devices. IoT middleware can support multiple radio access technologies and thus make integration of LPWA technologies with rest of IoT technologies straightforward. These middleware can also consolidate data from multiple sources to offer knowledge based value-added services to end-users. 

Interoperability is a still an open challenge. Testbeds and open-source tool chains for LPWA technologies are not yet widely available to evaluate interoperability mechanisms. 

%and require further advancement. 

%For applications requiring ultra-low latency, edge or fog computing can be considered to move processing closer to the access network. 

\subsection{Localization}
LPWA networks expect to generate significant revenue from logistics, supply chain management, and personal IoT applications, where location of mobile objects, vehicles, humans, and animals may be of utmost interest. An accurate localization support is thus an important feature for keeping track of valuables, kids, elderly, pets, shipments, vehicle fleets, etc. %Only a few LPWA networks offer such native support for localization. 
In fact, it is regarded as an important feature to enable new applications. 

Localization of mobile devices is typically achieved by properties of received signals~\cite{rssiranging} and time of flight based measurement. All such techniques require very accurate time synchronization and sufficient deployment density of base stations. This is rather easily achieved with a careful network deployment and planning. However, a very limited channel bandwidth of LPWA technologies and an often absence of a direct path between end devices and base stations introduce very large localization error~\cite{localizationerrors, chalmersthesis}. Thus, doing accurate localization using LPWA transceivers alone is a real challenge. 

%The localization accuracy is not reported from real practical settings. Therefore, experimental campaigns are required. As several other LPWA networks lack localization support,  
LPWA networks require new techniques that not only exploit physical layer properties~\cite{rssiranging} but also combine other established localization techniques to ascertain that accuracy is good enough for real tracking applications. 

%Experimental campaigns in real practical settings are also required to ascertain localization accuracy high enough for LPWA tracking applications. 

%\lora can localize the end-devices if its transmitted packet is received by multiple gateways.

%\subsection{Guaranteeing optimal data rates at different distances}
\subsection{Link optimizations and adaptability}

If a LPWA technology permits, each individual link should be optimized for high link quality and low energy consumption to maximize overall network capacity. Every LPWA technology allows multiple link level configurations that introduce tradeoffs between different performance metrics such as data rate, time-on-air, area coverage, etc. This motivates a need for adaptive techniques that can monitor link quality and then readjust its parameters for better performance. 

However for such techniques to work, a feedback from gateway to end devices is usually required over downlink. Link asymmetry that causes downlink of many LPWA technologies (e.g., \sigfox) to have a lower capacity than uplink is a major hurdle in this case and thus, needs to be addressed in some way. 

%Adaptive algorithms for adjusting data rate and RF power output to maximize throughput, battery lifetime and network capacity. Spreading factor for \lorawan. Duty cycle can be high if node does not use optimal modulation factors. For example nodes close to base station should have a possibility to send quickly and go back to sleep. 

%\subsection{Edge Computing techniques for IoTs}
%Data -> Knowledge Data compression
%Care application design to get

%\subsection{Densification Challenges arising from use of ISM band and Choice of modulation techniques}
%SLAs

\subsection{LPWA testbeds and tools}
LPWA technologies enable several smart city applications. A few smart city testbeds e.g.  SmartSantander\cite{smartsantander} have emerged in recent years. Such testbeds incorporate sensors equipped with different wireless technologies such as Wi-Fi, IEEE 802.15.4 based networks and cellular networks. However, there are so far no open testbeds for LPWA networks. Therefore, it is not cost-effective to widely design LPWA systems and compare their performance at a metropolitan scale. At the time of writing, only a handful of empirical studies~\cite{kartakis} compare two our more LPWA technologies under same conditions. In our opinion, it is a significant barrier to entry for potential customers. Providing LPWA technologies as a scientific instrumentation for general public through city governments can act as a confidence building measure. In the meanwhile, analytical models~\cite{trllorascale, randomnesscausingstrangeinterference} and simulators~\cite{lorasim, padovathesis} have recently been proposed for the popular LPWA technologies.  

% Some standards are even more closed than others, e.g., Sigfox! Unique 

%\subsection{Compliance to Regional Regulations on Spectrum use}
%Seamless International roaming: cost of devices to support multiple bands probably not acceptable. 
%
%Extending business across the globe
%
%Assuring the same performance. 
%
%Regulatory barriers. 
%challenge would be to guarantee reliability and security. 

\subsection{Authentication, Security, and Privacy}

Authentication, security, and privacy are some of the most important features of any communication system. Cellular networks provide proven authentication, security, and privacy mechanisms. Use of Subscriber Identity Modules (SIM) simplifies identification and authentication of the cellular devices. LPWA technologies, due to their cost and energy considerations, not only settle for simpler communication protocols but also depart from SIM based authentication. Techniques and protocols are thus required to provide equivalent or better authentication support for LPWA technologies. Further to assure that end devices are not exposed to any security risks over prolonged duration, a support for over-the-air (OTA) updates is a crucial feature. A lack of adequate support for OTA updates poses a great security risk to most LPWA technologies. 

%The ability to update the software running on the end-devices is required. Asymmetric communication pose a challenge. Support for bi-directional 

% and need more innovative solutions........

Margelis et al.~\cite{bristol} highlight a few security vulnerabilities of the three prominent LPWA technologies namely \sigfox, \lorawan, and \ingenu. To offer an example, end devices in  \sigfox and \lorawan networks do not encrypt application payload and the network join request respectively~\cite{bristol}, potentially leading to eavesdropping. Further most LPWA technologies use symmetric key cryptography in which the end devices and the networks share a same secret key. 
Robust and low-power mechanisms for authentication, security, and privacy need further investigation.  

%\sigfox leave encryption of payload to application. (native encryption support for application data - none)
%authentication secret key does not change. 
%blacklisting of node
%
%\lora: nonce generated by multiple measurements of RSSI, which may not be \emph{uniformly} random 
%\usman{patent idea related to \lora over the air authentication: randomizing nonces (send at multiple SFs) to generate  uniformly random RSSI measurements from sequence of measurements }
%How to generate network and session keys not mentioned leaving responsibility to application designers. 
%absence of CRC in downlink communication --> message integrity
%
%
%
%\ingenu \cite{bristol} all devices use same code to spread the signals. An attacker can join the network to get access to this shared code and then eavesdrop traffic of end-devices. 
%man-in-the-middle attack. 
%
%and then compromise potential security of all 
%\ingenu: all nodes sharing the same spreading code --> eavesdropping, traffic analysis
%impersonation as a base station: grabbing the join requests to get the get the spreading code

\subsection{Mobility and Roaming}

%Unlicensed ISM band has typically been used for short range technologies such as Wifi.  Because of their limited coverage, they do not provide support for roaming and mobility by design. 

Roaming of devices between different network operators is a vital feature responsible for the commercial success of cellular networks. 
%Most LPWA networks, however, do not provide roaming support at the time of writing. 
Whilst some LPWA technologies do not have the notion of roaming (work on a global scale such as \sigfox), there are others that do not have support for roaming as of the time of this writing. The major challenge is to provide roaming without compromising the lifetime of the devices. To this effect, the roaming support should put minimal burden on the  battery powered end-devices.  Because the end-devices duty cycle aggressively, it is reasonable to assume that the low power devices cannot receive downlink traffic at all times. Data exchanges over the uplink should be exploited more aggressively. Network assignment is to be resolved in backend systems as opposed to the access network. All the issues related to agility of roaming process and efficient resource management have to be addressed. 

%Under the assumptions stated above, several base stations should promiscuously listen to the end-devices, even if the devices are not assigned to these networks. It introduces a high communication overhead in the access network as well as between networks servers. 
Further billing and revenue sharing models for roaming across different networks have to be agreed upon. 

International roaming across regions controlled by different spectrum regulations (e.g., USA, Europe or China) is even more challenging. In order to comply to varying spectrum regulations, end devices should be equipped with capabilities to detect the region first and then adhere to the appropriate regional requirements when transmitting data. This adds complexity to end devices and therefore the cost. Simple low cost design to support international roaming is thus required.  

% will be even more challenging because all use different channels. There is a need to 
%Devices operating in ISM bands will need to detect the region first and comply to varying regulations. This adds significant complexity to the end user.  

\subsection{Support for Service Level Agreements}
The ability to offer certain QoS guarantees can be a competitive differentiator between different LPWA operators. While it is relatively easy to offer QoS guarantees in the licensed spectrum, most proprietary technologies opt for the license-exempt spectrum for a faster time to market. As a result, they have to adhere to regional regulations on the use of shared spectrum, which may limit the radio duty cycle and transmitted RF power. Cross-technology interference also influences the performance of LPWA technologies. 
%Vital to keep the connectivity cost for low-end IoT devices low.   restricts the scalability of the networks. 

Providing carrier grade performance on a spectrum shared across multiple uncoordinated technologies and tens of thousands of devices per base station is a significant challenge. Service Level Agreements (SLAs) are likely to be violated due to the factors outside the control of network operators. Therefore, the support for SLAs is expected to be limited in license-exempt bands. Studying such extremely noisy environments to know if some relaxed statistical service guarantees can be provided is a good potential research direction. %Furthermore, systems have put in place to monitor the compliance of SLAs.   

%Other factors such as link asymmetry cause longer delays over downlink  
%
%Not very easy to get: Number of contending devices. Need for adaptive techniques to assure an ultra-low latency for control-loop industrial applications (especially in downlink communication).
%Low latency for closing the loop in industrial applications.
%
%Sigfox uses software defined radios in the basestations . ...  

%Typically LPWA are addressing low data rate applications that can tolerate some communication latency. However   

\subsection{Co-existence of LPWA technologies with other wireless networks}

Each application has a unique set of requirements, which may vary over different time scales and contexts. If connectivity of the end-devices is supplemented with LPWA technologies in addition to the cellular or wireless LANs, operation of applications can be optimized. Conflicting goals like energy efficiency, high throughput, ultra-low latency and wide area coverage can be achieved by leveraging the benefits of each technology~\cite{ahlpwa, ltelpwa}. System-level research is needed to explore benefits of such opportunistic and contextual network access. 

%As future LPWA as well as conventional cellular networks are managed by MNOs, use cases can be envisages in which both networks cooperate to optimize the each operation for energy, wider coverage etc.

There can be different use cases where multiple technologies can cooperate with each other. The ETSI LTN specification~\cite{ltn1} lists a few of these use cases for cellular/LPWA cooperation. To offer an example, when cellular connectivity is not available, LPWA technologies can still be used as a fall-back option for sending only low data rate critical traffic. Further, the periodic keep-alive messages of cellular networks can be delegated to energy-efficient LPWA networks~\cite{ltn1}. % %Availability of geo-location service in some of LPWA networks is very convenient. 
There can be other novel ways for cooperation between LPWA and cellular networks. For instance, LPWA technologies can assist route formation for the device-to-device communication in cellular networks. When some devices outside the cellular coverage need to build a multi-hop route to reach cellular infrastructure, LPWA connectivity can assist in detecting proximity to other serviced devices. These use-cases may have a strong appeal for public safety applications. Further, as we know, LPWA technologies are designed specifically for ultra low data rates. A need of occasionally sending large traffic volumes can be met with a complementary cellular connection, which can be activated only on demand. 

A joint ownership of LPWA and cellular networks combined with a drop in prices of LPWA devices and connectivity make a strong business case for the above-mentioned use cases. However, there is a need to overcome many systems related challenges. 

%and  define interfaces between LPWA and other wireless networks and . 
 
% Low cost of LPWA end devices
%---> use cases in which these networks cooperate

%signaling can be 
%D2D communication 

%To enable such cooperative communication, 
\input{business_table}

\subsection{Support for Data Analytics}

Compared to a human subscriber, the average revenue generated by a single connected M2M/IoT device is rather small. Therefore, network operators see a clear incentive in extending their business beyond the pure connectivity for sake of a higher profitability. One way to do so is by augmenting LPWA networks with sophisticated data analytics support that can convert the raw collected data into contextually relevant information for the end-users. Such knowledge can support end users in making intelligent decisions, earning higher profits, or bringing their operational costs down. Network operators thus can monetize this by selling knowledge to end users. 

There are however enormous challenges associated with providing a LPWA network as a service to the end-users. It requires a unified management of business platform and a scalable integration with the cloud. One of the main challenges is also to offer custom-tailored services to many different vertical industries, effectively covering different use cases ideally by a single LPWA technology.

%lack of experience in content provider 
%MNOs may like to 
%Early adopter analysis and profitability aspects. 

%\subsection{Uncertainty with Longevity of the Solution}
%\usman{section assigned to parag}
%Switching a provider and even technology

%*****************************************************************************
\input{business}

%*****************************************************************************
\section{Conclusion}
\label{sec:conclusion}
Wide area coverage, low power consumption, and inexpensive wireless connectivity blends together in LPWA technologies to enable a strong business case for low throughput IoT/M2M applications that do not require ultra-low latency. However, this combination of often conflicting goals is a result of carefully designed physical and MAC layer techniques, precisely surveyed in this paper. To tap into the huge IoT/M2M market, several commercial providers exploit different innovative techniques in their LPWA connectivity solutions. The variety of these solutions have resulted in a fragmented market, highlighting a dire need for standards. We provided a comprehensive overview of many such standardization efforts led by several SDOs and SIGs. We observe that most standards focus on physical and MAC layers. A gap at the upper layers (application, transport, network etc.) is to be bridged. Further, we point out important challenges that LPWA technologies face today and possible directions to overcome them. We encourage further developments in LPWA technologies to push the envelop of connecting massive number of devices in future.  

%We have precisely surveyed different commercial solutions, their techniques and how these enable 
 
%Several standards developing organizations are engaged in proposing standards for LPWA, with major focus on PHY and MAC layer than upper layers. 
%
%While these technologies are already hitting the market with slow. 
%long-term 
%severl challenges are yet to be addressed. 

%LPWA specifically address low power devices. 
%less silicon running lean protocols
%While there is a need to put more intelligence in the network to provide support of mobility, roaming, scalable operation and spectral efficiency.  
%such intelligence can be provided by cognitive software defined radios in access networks, 
%cognitive software defined radios, cloud computing, virtualization

%\bibliographystyle{ieeetr}
%\bibliography{library}

\balance
\bibliographystyle{IEEEtran}
\bibliography{./paper}
\end{document}

%% file: macros.tex
\newboolean{authnotes}
\setboolean{authnotes}{false}
\ifthenelse{\boolean{authnotes}}
{
\newcommand{\memo}[1]{{\color{red} #1}}
\newcommand{\usman}[1]{}
\newcommand{\parag}[1]{}
\usepackage{draftwatermark}
\SetWatermarkText{}
}
{
\newcommand{\usman}[1]{}
\newcommand{\parag}[1]{}
\usepackage{draftwatermark}
\SetWatermarkText{}
\newcommand{\memo}[1]{#1}
}

\newcommand{\fakeparagraph}[1]{\vspace{1mm}\noindent\textbf{#1}}

\newcommand{\cmark}{\ \includegraphics[width=0.2cm]{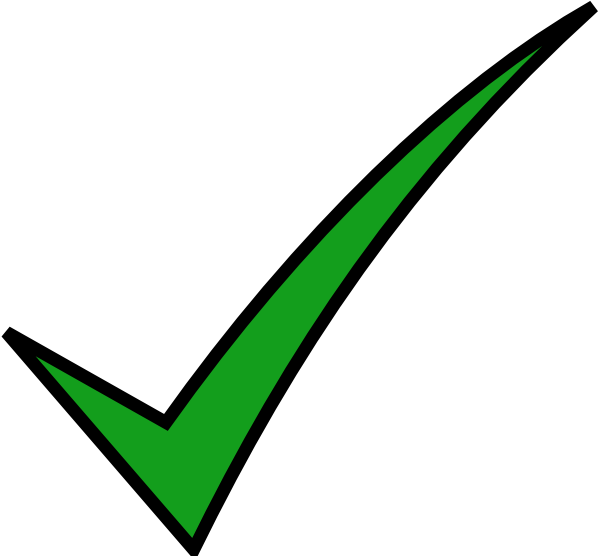}}%
\newcommand{\xmark}{ \includegraphics[width=0.2cm]{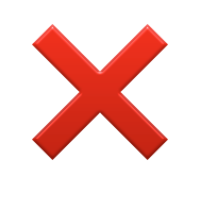}}%

\newcommand{\lora}{\textsc{LoR}a\xspace}
\newcommand{\lorawan}{\textsc{LoRaWAN}\xspace}
\newcommand{\sigfox}{\textsc{SigFox}\xspace}
\newcommand{\ingenu}{\textsc{Ingenu}\xspace}
\newcommand{\rpma}{\textsc{RPMA}\xspace}
\newcommand{\weightless}{\textsc{Weightless}\xspace}
\newcommand{\qowisio}{\textsc{Qowisio}\xspace}
\newcommand{\dash}{\textsc{Dash7}\xspace}
\newcommand{\telensa}{\textsc{Telensa}\xspace}
\newcommand{\nbiot}{\textsc{NB-I}o\textsc{T}\xspace}

\newcommand{\aloha}{\textsc{Aloha}\xspace}

\newcommand{\lorawantm}{\textsc{LoR}aWAN\texttrademark\xspace} 

\newcommand{\loraalliance}{\textsc{LoR}a\texttrademark\xspace Alliance\xspace} 

\newcommand{\ul}{\textsc{(ul)}\xspace}
\newcommand{\dl}{\textsc{(dl)}\xspace}

\newcommand{\urban}{\textsc{(urban)}\xspace}
\newcommand{\rural}{\textsc{(rural)}\xspace}

\newcommand{\ism}{\textsc{ISM}\xspace}

\newcommand{\subghz}{\textsc{Sub-GHz}\xspace}

%% file: abstract.tex
\begin{abstract}
Low Power Wide Area (LPWA) networks are attracting a lot of attention primarily because of their ability to offer affordable connectivity to the low-power devices distributed over very large geographical areas. In realizing the vision of the Internet of Things (IoT), LPWA technologies complement and sometimes supersede the conventional cellular and short range wireless technologies in performance for various emerging smart city and machine-to-machine (M2M) applications. This review paper presents the design goals and the techniques, which different LPWA technologies exploit to offer wide-area coverage to low-power devices at the expense of low data rates. We survey several emerging LPWA technologies and the standardization activities carried out by different standards development organizations (e.g., IEEE, IETF, 3GPP, ETSI) as well as the industrial consortia built around individual LPWA technologies (e.g., \loraalliance, \weightless-SIG, and \dash Alliance). We further note that LPWA technologies adopt similar approaches, thus sharing similar limitations and challenges. This paper expands on these research challenges and identifies potential directions to address them. While the proprietary LPWA technologies are already hitting the market with large nationwide roll-outs, this paper encourages an active engagement of the research community in solving problems that will shape the connectivity of tens of billions of devices in the next decade. 
\end{abstract}

%% file: intro.tex
\section{Introduction}
\label{sec:intro}

%MARKET FACTORS CREATING BUISNESS OPPORTUNITIES
%--BUSINESS FORECASTS
%--DECOMMISSIONING OF 2G, 3G
%--2G WERE NOT AS ENERGY EFFICIENT AS NEW TECHNOLOGIES 
%ALL THESE THREE FACTORS 
%Support with Surveys, Statistics on market growth??? Industrial partner facing sheer competition.~\cite{polastre04:versatile}

\IEEEPARstart{T}{he} Internet of Things (IoT) promises to revolutionize %our everyday life, industrial production, and business processes. 
the way we live and work.
It could help us in overcoming the top global challenges due to population explosion, energy crisis, resource depletion, and environmental pollution. To realize this vision, \emph{things} need to sense their environment, share this information among themselves as well as with humans to enable intelligent decision making for positively affecting our entire ecosystem. Due to this promise, an interest in IoT is phenomenal. Multiple independent studies have forecasted a rampant growth in volume and revenue of IoT and Machine-to-Machine (M2M) industry in the next ten years. Number of connected M2M devices and consumer electronics will surpass the number of human subscribers using mobile phones, personal computers, laptops and tablets by 2020~\cite{ericssonwhitepaper}. Moving forward, by 2024, the overall IoT industry is expected to generate a revenue of 4.3 trillion dollars\cite{machina15} across different sectors such as device manufacturing, connectivity, and other value added services. Recent improvements in cheap sensor and actuation technologies along with an emergence of novel communication technologies are all positive indicators, supporting the forecasted trends. 

\begin{figure*}
\includegraphics[scale=0.75]{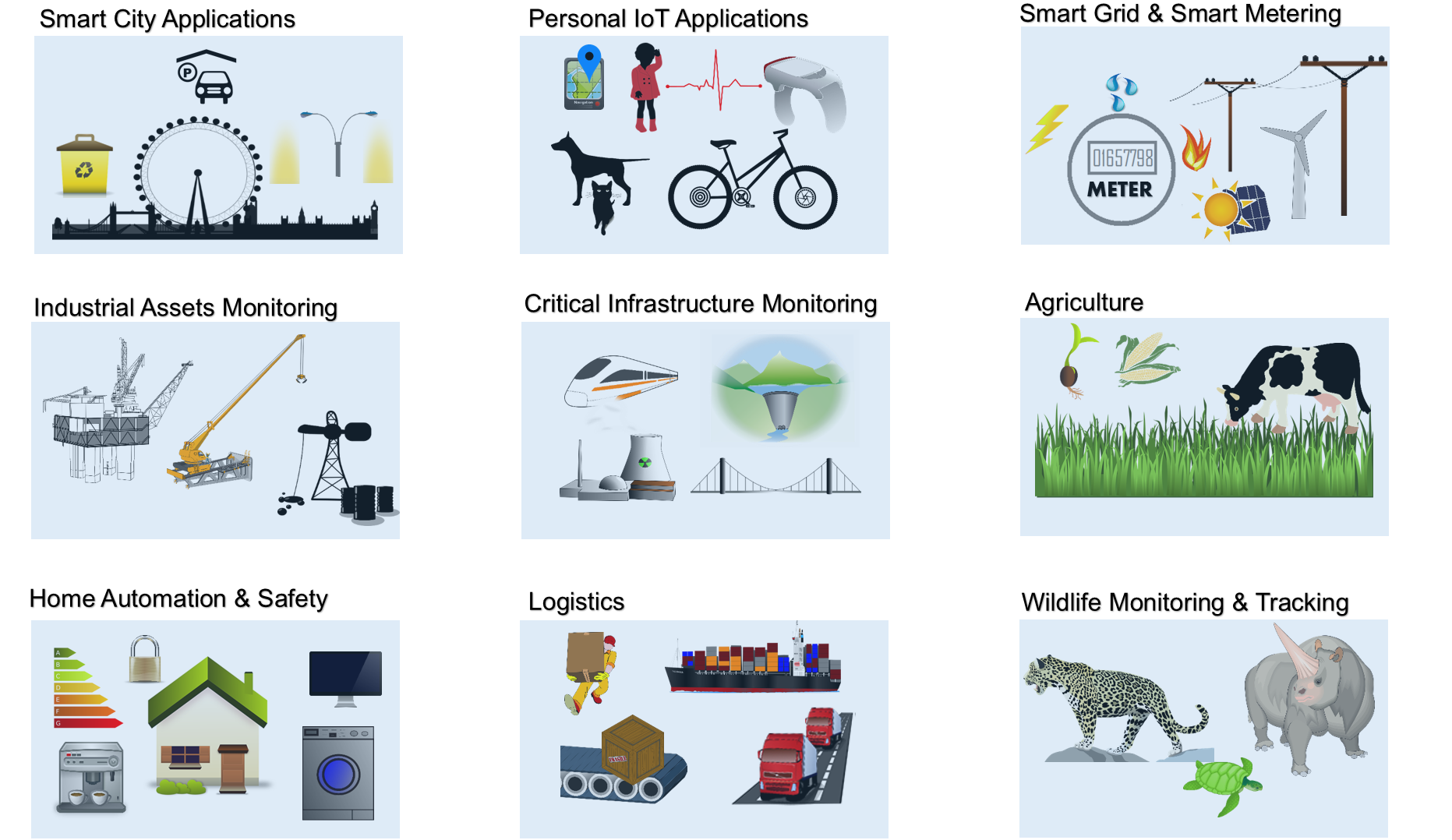}
\caption{Applications of LPWA technologies across different sectors}
\label{fig:applications}
\end{figure*}

Low Power Wide Area (LPWA) networks represent a novel communication paradigm, which will complement traditional cellular and short range wireless technologies in addressing diverse requirements of IoT applications. LPWA technologies offer unique sets of features including wide-area connectivity for low power and low data rate devices, not provided by legacy wireless technologies. Their market is expected to be huge. Approximately one fourth of overall 30 billion IoT/M2M devices are to be connected to the Internet using LPWA networks using either proprietary or cellular technologies\memo{~\cite{nokiawhitepaper}}. Figure~\ref{fig:applications} highlights variety of applications across several business sectors that can exploit LPWA technologies to connect their end devices. These business sectors include but not limited to smart city, personal IoT applications, smart grid, smart metering, logistics, industrial monitoring, agriculture, etc.

%An abundance of these devices in near future requires availability of cheap sensors, actuators and innovative communication technologies to offer ubiquitous coverage of millions of devices over long geographical distances.  

%However, cheap ubiquitous connectivity of these devices, which can be distributed over very large geographical distances, is a key enabler of this vision. Many of these devices can be distributed over large geographical distances and send infrequent data such as in smart city scenarios. 

%While availability of low cost  availability of low cost devices equipped with sensing, processing and communication capabilities. step in the right direction.  is key to realise several smart city applications.  

%new breed of applications in smart city, personal health care where  

%place and play  
%SMART CITY

%define the scope of the paper in a wider landscape of IoT technologies.
%short range wireless, cellular, unlicensed/licencesd LPWAN

%In contrast to traditional wireless technologies, LPWA networks make different trade-offs among wide coverage, power consumption and the throughput and therefore presents a unique set of features required by different IoT applications. 

LPWA networks are unique because they make different tradeoffs than the traditional technologies prevalent in IoT landscape such as short-range wireless networks e.g., ZigBee, Bluetooth, Z-Wave, \memo{legacy} wireless local area networks (WLANs) e.g., Wi-Fi,  and cellular networks e.g. Global System for Mobile Communications (GSM), Long-Term Evolution (LTE) etc. 
%Figure~\ref{fig:lpwvsothers} highlights these differences. 
The legacy non-cellular wireless technologies are not ideal to connect low power devices distributed over large geographical areas. The range of these technologies is limited to a few hundred meters at best. The devices, therefore, cannot be arbitrarily deployed or moved \emph{anywhere}, a requirement for many applications for smart city, logistics and personal health~\cite{techniquesprototypes}. The range of these technologies is extended using a dense deployment of devices and gateways connected using multihop mesh networking. Large deployments are thus prohibitively expensive. \memo{Legacy} WLANs, on the other hand, are characterized by shorter coverage areas and higher power consumption for machine-type communication (MTC). 

%\begin{figure}
%\includegraphics[scale=0.4]{img/technology_tradeoffs-better.pdf}
%\caption{LPWA vs. legacy wireless technologies}
%\label{fig:lpwvsothers}
%\end{figure}

%First,  Second, devices extend their range by communicating over mutlihop mesh networks. Because the traffic load varies significantly across the mesh network, some devices deplete their battery faster than others and therefore do not meet a 10 year battery lifetime requirement. Third, short range wireless technologies require a dense deployment of devices as well as gateways to connect to the Internet, increasing the investment cost on infrastructure.
%Third, a higher datarate and low receiver sensitivity results in a link budget, low enough to reach long distances or penetrate obstracles efficiently.   
%Last 15 years more research on Low power personal area networks. Low power deployments limited to buildings, tunnels etc. 
%However various smart city applications require coordination between geographically distant entities at a city scale.

A wide area coverage is provided by cellular networks, a reason of a wide adoption of second generation (2G) and third generation (3G) technologies for M2M communication. However, an impending decommissioning of these technologies~\cite{decommission3g}, as announced by some mobile network operators (MNOs), will broaden the technology gap in connecting low-power devices. In general, traditional cellular technologies do not achieve energy efficiency high enough to offer ten years of battery lifetime. The complexity and cost of cellular devices is high due to their ability to deal with complex waveforms, optimized for voice, high speed data services, and text. For low-power MTC, there is a clear need to strip complexity to reduce cost. Efforts in this direction are underway for cellular networks by the Third Generation Partnership Project and are covered as part of the discussion in Section~\ref{sec:standards}. 

%An increased interest in LPWA also comes from recent announcements fom operators about decommissioning of 2G cellular networks in future. Several M2M application that currently rely on 2G will be left with no other choice than to switch to an alternative and better connectivity technology.
% 
With a phenomenal range of a few to tens of kilometers~\cite{lpwanloracoverage} and battery life of ten years and beyond, LPWA technologies are promising for the Internet of low-power, low-cost, and low-throughput things. A very long range of LPWA technologies enables devices to spread and move over large geographical areas. IoT and M2M devices connected by LPWA technologies can be turned on \emph{anywhere and anytime} to sense and interact with their environment instantly. It is worth clarifying that LPWA technologies achieve long range and low power operation at the expense of low data rate (typically in orders of tens of kilobits per seconds) and higher latency (typically in orders of seconds or minutes). %These technologies do not cater to high throughput applications requiring ultra-low latency such as wireless industrial control. 
\memo{Therefore it is clear that LPWA technologies are not meant to address each and every IoT use case and caters to a niche area in IoT landscape. Specifically, LPWA technologies are considered for those use cases that are delay tolerant, do not need high data rates, and typically require low power consumption and low cost, the latter being an important aspect. Such MTC application are categorized as Massive MTC~\cite{5gradioaccess} in contrast to Critical MTC~\cite{5gradioaccess} applications that require ultra-low latency and ultra high reliability. The latter are definitely out of the remit of LPWA technologies because their stringent performance requirements such as up to five nines (99.999\%) reliability and up to 1-10 ms latency cannot be guaranteed with a low cost and low power solution. While LPWA technologies, for this reason,  are not suitable for many industrial IoT, vehicle to vehicle (V2V), and vehicle to infrastructure (V2I) applications~\cite{understandinglimitsoflora}, they still meet the needs of a plethora of applications for smart cities, smart metering, home automation, wearable electronics, logistics, environmental monitoring etc. (see Figure~\ref{fig:applications}) that exchange small amount of data and that also infrequently.} Therefore, appeal of LPWA technologies, although limited by its low data rate, is still broad. This is the reason why LPWA technologies generated so much interest after the proprietary technologies such as \sigfox~\cite{sigfox} and \lora~\cite{lorawan} hit the market. 

At this moment, there are several competing LPWA technologies, each employing various techniques to achieve long range, low power operation, and high scalability. Section~\ref{sec:goals_techniques} presents these design goals and describes how a combination of different novel techniques actually achieves them. Section~\ref{sec:proprietary} then discusses several early proprietary LPWA technologies and their technical features, highlighting the need for standardization to flourish IoT ecosystem. %, pros and cons \usman{history? read again}. 
%With LPWA technologies evolving over several fronts, the standardization is a key enabler for a flourishing IoT ecosystem. 
%LPWA are constantly evolving on several fronts. First, 
To this effect, several well-known standard developing organizations (SDOs) such as European Telecommunications Standard Institute (ETSI)~\cite{etsi}, Third Generation Partnership Project (3GPP)~\cite{3gpp}, Institute of Electrical and Electronics Engineers (IEEE)~\cite{ieee}, and Internet Engineering Task Force (IETF)~\cite{ietf} are working towards the open standards for LPWA technologies. Further, multiple industrial alliances are built around individual LPWA technologies to promote new standards. \loraalliance~\cite{loraalliance}, \weightless-SIG~\cite{weightless} and \dash Alliance~\cite{dash7alliance} are a few examples of such special interest groups (SIGs). Section~\ref{sec:standards} covers the standardization efforts led by all these SDOs and SIGs.  %and their future planned releases\usman{rephrase}. 

On a technical side, LPWA providers need to push innovative solutions to overcome the challenge of connecting massive number of IoT and M2M devices. It is indeed not an easy task especially when the heterogeneous LPWA technologies share limited radio resources to render scalable and secure connectivity to low-power and inexpensive end devices. 
%especially because most proprietary solutions operate in the unlicensed Industrial, Scientific and Medical (ISM) radio band, primarily to expedite delivery of their services to the market and avoid exorbitant cost of licensed spectrum. However, interference in this shared band is a major challenge for co-existence of LPWA with other technologies.
Multiple trade-offs made by the LPWA technologies bring several challenges, which are discussed in Section~\ref{sec:challenges} along with possible research directions to address them. Section~\ref{sec:business} then highlights business considerations for LPWA technologies before finally concluding this paper.

%CHECK ITU -- LPWA STANDARD? 

%DRAW SOME SPIDER CHARTS.
% Describe the contents of the paper. 
%use-cases 
%techniques 
%LPWA technologies and taxonomy 
%challenges main contribution. 
%business model 
%
%sensors processors cost size energy consumption reducing 
%communication has been bottleneck.  
%
%Figure showing how LPWA are different than rest of communication technologies such as cellular, PAN, wifi, cellular in terms of range power consumption and system throughput.  
%
%place and play requirement. 
%easy integration 
%cloud 
%
%transition from M2M to Internet of Things
%M2M will make a transition from 2G to LPWA. Cost is expected to go down and so new use cases will emerge.  
%
% 
%Cheap end-devices. 
%High sensitivity modulation techniques. Low infrastructure cost. 
%A data rate sufficient for diverse 
%Taxonomy 
%Less disruptive (Evolution) --- more disruptive (Completely new infrastructure)
%Evolution of existing as well as completely new technologies.
%Modulation techniques used
%Band used 
%Choice of band to some extends define how much interference will be experienced: ISM vs. licensed band 

%*****************************************************************************

%*****************************************************************************

%% file: goals_techniques.tex
\section{Design Goals and Techniques}
\label{sec:goals_techniques}

%This section describes the key design goals for LPWA and the techniques to achieve them. 
The success of LPWA technologies lies in their ability to offer low-power connectivity to massive number of devices distributed over large geographical areas at an unprecedented low-cost.  This section describes the techniques LPWA technologies used to achieve these often conflicting goals. \memo{ We like to highlight that LPWA technologies share some of the design goals with other wireless technologies. The key objective of LPWA technologies is, however, to achieve a long range with low power consumption and low cost unlike that of the other technologies for which achieving higher data rate, lower latency and higher reliability may be more important.}

\subsection{Long range}

LPWA technologies are designed for a wide area coverage and an excellent signal propagation to hard-to-reach indoor places such as basements. Quantitatively, a +20 dB gain over legacy cellular systems is targeted. This allows the end-devices to connect to the base stations at a distance ranging from a few to tens of kilometers depending on their deployment environment (rural, urban, etc.). Sub-GHz band and special modulation schemes, discussed next, are exploited to achieve this goal. 

%high link budget not only extends the coverage to rural areas but also improves signal penetration in hard-to-reach indoor places 
%+20 dB gain over legacy GPRS
%higher receiver sensitivity. 
 %enhanced coverage,
%Better indoor coverage
%better penetration 
%hard to reach places meters

\subsubsection{Use of Sub-1GHz band}
With an exception of a few LPWA technologies (e.g., \weightless-W~\cite{weightless} and \ingenu~\cite{howrpmaworks}), most use Sub-GHz band, which offers robust and reliable communication at low power budgets. Firstly, compared to the 2.4 GHz band, the lower frequency signals experience less attenuation and multipath fading caused by obstacles and dense surfaces like concrete walls. Secondly, sub-GHz is less congested than 2.4 GHz, a band used by most-popular wireless technologies e.g., Wi-Fi, cordless phones, Bluetooth, ZigBee, and other home appliances. The resulting higher reliability enables long range and low power communication. Nevertheless, the \ingenu's RPMA technology ~\cite{howrpmaworks} is an exception that still exploits 2.4 GHz band due to more relaxed spectrum regulations on radio duty cycle and maximum transmission power in this band across multiple regions.

\subsubsection{Modulation Techniques}
LPWA technologies are designed to achieve a link budget of $150 \pm 10$ dB that enables a range of a few kilometers and tens of kilometers in urban and rural areas respectively. The physical layer compromises on high data rate and slows downs the modulation rate to put more energy in each transmitted bit (or symbol). Due to this reason, the receivers can decode severely attenuated signals correctly. Typical sensitivity of state of the art LPWA receivers reaches as low as -130 dBm. 
%A way to increase sensitivity is to increase energy per symbol or bit, which can be done by slowing down the modulation rate. It increase the range at an expense of lower data rate. 
Two classes of modulation techniques namely narrowband and spread spectrum techniques have been adopted by different LPWA technologies. 

\fakeparagraph{Narrowband} modulation techniques provide a high link budget by encoding the signal in low bandwidth (usually less than 25kHz). By assigning each carrier a very narrow band, these modulation techniques share the overall spectrum very efficiently between multiple links. The noise level experienced inside a single narrowband is also minimal. Therefore, no \memo{processing gain through frequency de-spreading} is required to decode the signal at the receiver, resulting in simple and inexpensive transceiver design. \nbiot and \weightless-P are examples of narrowband technologies. 

A few LPWA technologies squeeze each carrier signal in an \emph{ultra} narrow band (UNB) of width as short as 100Hz (e.g., in \sigfox), further reducing the experienced noise and increasing the number of supported end-devices per unit bandwidth. However, the effective data rate for individual end devices decreases as well, thus increasing the amount of time the radio needs to be kept ON. 
%Reduced data rate increases the radio on time for sending packets, which is  legislated to 
%When sharing license-exempt \ism band, the regional regulations limit transmit power or radio duty cycle. 
This low data rate in combination with spectrum regulations on sharing underlying bands may limit maximum size and transmission frequency of data packets, limiting number of business use cases. \sigfox, \weightless-N and \telensa~\cite{telensapatent} are a few examples of LPWA technologies that use UNB modulation.

\usman{A few LPWA technologies like \sigfox and \telensa use ultra narrow band. Weightless-N uses differential binary phase shift keying (DBPSK)}
 
\fakeparagraph{Spread spectrum techniques} 
%spreads a narrowband signal over large frequency. Multiple orthoganl signals are however can be transmitted over the same wideband to increase the overall network capacity.  
spread a narrowband signal over a wider frequency band but with the same power density\usman{check again}. The actual transmission is a noise-like signal that is harder to detect by an eavesdropper, more resilient to interference, and robust to jamming attacks. More processing gain is however required on the receiver side to decode the signal that is typically received below the noise floor. Spreading a narrowband signal over a wide band results in less efficient use of the spectrum. But, this problem is typically overcome by the use of multiple orthogonal sequences. %or spreading factors. 
\memo{ As long as multiple end devices use different channels and/or orthogonal sequences, all can be decoded concurrently, resulting in a higher overall network capacity}. Different \memo{variants} of spread spectrum techniques are used by existing standards \memo{as discussed in Section~\ref{subsec:lora} and Section~\ref{subsec:rpma}}. Chirp Spread Spectrum (CSS) and Direct Sequence Spread Spectrum (DSSS) are used by \lora and \rpma respectively.  

\usman{error correction?? FEC etc?}

\subsection{Ultra low power operation}
\label{sec:lowpower}
Ulra-low power operation is a key requirement to tap into the huge business opportunity provided by battery-powered IoT/M2M devices. A battery lifetime of 10 years or more with AA or coin cell batteries is desirable to bring the maintenance cost down.  

\subsubsection{Topology}

While mesh topology has been extensively used to extend the coverage of short range wireless networks, their high deployment cost is a major disadvantage in connecting large number of geographically distributed devices. %The devices cannot be arbitrary placed anywhere in large geographical areas like cities. 
%However this is not the only problem when connecting battery powered devices. 
Further, as the traffic is forwarded over multiple hops towards a gateway, some nodes get more congested than others depending on their location or network traffic patterns. Therefore, they deplete their batteries quickly, limiting overall network lifetime to only a few months to years~\cite{jsan2030509, Oppermann2014}.

On the other hand, a very long range of LPWA technologies overcomes these limitations by connecting end devices \emph{directly} to base stations, obviating the need for the dense and expensive deployments of relays and gateways altogether. The resulting topology is a star that is used extensively in cellular networks and brings huge energy saving advantages. As opposed to the mesh topology, the devices need not to waste precious energy in busy-listening to other devices that want to relay their traffic through them. An always-on base station provides convenient and quick access when required by the end-devices. 

In addition to star, a few LPWA technologies support tree and mesh topologies but with extra complexity in protocol design. 

\subsubsection{Duty Cycling}
Low power operation is achieved by opportunistically turning off power hungry components of M2M/IoT devices~\cite{demirkol2006mac, anastasi2009energy} e.g., data transceiver. Radio duty cycling allows LPWA end devices to turn off their transceivers, when not required. Only when the data is to be transmitted or received, the transceiver is turned on. 

LPWA duty cycling mechanisms are adapted based on application, type of power source, and traffic pattern among other factors. If an application needs to transfer the data only over the uplink, the end devices may wakeup only when data is ready to be transmitted. In contrast, if downlink transmissions are required as well, the end devices make sure to listen when the base station actually transmits. The end devices achieve this by agreeing on a listening schedule. For example, the end devices may listen for a short duration after their uplink transmissions to receive a reply back. Alternatively, they may wakeup at a scheduled time agreed with the base station. For main-powered end devices requiring an ultra-low latency downlink communication, radio transceiver can stay in an always on mode. Different LPWA standards such as \lorawan~\cite{lorawan} define multiple classes of the end devices based on their communication needs in uplink or downlink. 

In realm of LPWA technologies, duty cycling the data transceiver is not only a power saving mechanism but also a legislative requirement. Regional regulations on sharing spectrum~\cite{spectrumuse} may limit the time a single transmitter can occupy to assure its coexistence with other devices sharing the same channel\usman{should I give an example from FCC or EU?}.  

Duty cycling can also be extended beyond the transceiver to other hardware components, as explored in the context of many low-power embedded networks~\cite{1632657, Raza:2016:TAE:2938783.2938873}. Modular hardware design may provide ability to choose different operational modes and turn on or off individual hardware components (such as auxiliary components and storage and micro-controllers)~\cite{raza2015energy}. By exploiting these power management techniques, LPWA application developers can further reduce the power consumption and increase the battery lifetime. %Depending on application requirements, They can also trade-off energy efficiency for increased application performance such as quality of sensed phenomenon. 
%All these mechanisms are 

\subsubsection{Lightweight Medium Access Control}
%Low-power operation asks for 
Most-widely used Medium Access Control (MAC) protocols for cellular networks or short range wireless networks are too complex for LPWA technologies. For example, cellular networks synchronize the base stations and the user equipment (UE) accurately to benefit from complex MAC schemes that exploit frequency and time diversity. The control overhead of these schemes, while justifiable for powerful cellular UEs, is substantial for the LPWA end devices. Put differently, the control of these MAC protocols may be even more expensive than the short and infrequent machine type communication\usman{give number?} of LPWA devices. Further, a very tight synchronization needed by these schemes is difficult to be met by ultra low-cost (\$1-\$5) end devices having low quality cheap oscillators. When accessing the spectrum, these devices experience drift in both time and frequency domains, making an exclusive access to the shared medium a primary challenge for the competing devices. Due to this reason, simple random access schemes are more popular for LPWA technologies.\usman{is it a GENERALIZATION!}
%unpredictable traffic patterns

Carrier sense multiple access with collision avoidance (CSMA/CA) is one of the most popular MAC protocols successfully deployed in WLANs and other short range wireless networks. The number of devices per base-station are limited for such networks, keeping the hidden node problem at bay. However, as the number of these devices grow in LPWA networks, carrier sensing becomes less effective and expensive~\cite{csexpensive} in reliably detecting on-going transmissions, negatively affecting the network performance. While virtual carrier sensing using Request to Send/ Clear to Send (RTS/CTS) mechanism is used to overcome this problem, it introduces extra communication overhead over the uplink and the downlink. With massive number of devices, LPWA technologies cannot usually afford this excessive signaling overhead. In addition, link asymmetry, a property of many LPWA technologies today, reduces the practicality of virtual carrier sensing. \usman{Add? CSMA/CA not always possible: long-distance signal is too weak }. 

Due to these reasons, multiple LPWA technologies such as \sigfox and \lorawan resort to the use of \aloha, a random access MAC protocol in which end devices transmit without doing any carrier sensing. The simplicity of \aloha is thought to keep design of transceiver simple and low cost\usman{put future reference to why \aloha works}. \memo{Nevertheless, TDMA based MAC protocols are also considered by \ingenu and \nbiot to allocate radio resources more efficiently although at the expense of more complexity and cost for end devices.}

%Several technologies such as \sigfox and \lorawan use \aloha to keep MAC protocol design simple and therefore less power hungry. 
%Don’t listen just transmit (SIGFOX, LoRa)
%But why does Aloha even work? 
%PHY layer rescues!
%Multiple signals overlapping in time can be decoded as they are orthogonal to each other e.g., use of multiple spreading factors
%MAC protocol 
%random access techniques like \aloha

%\subsubsection{Offloading Intelligence from the Access Network to the Backhaul}
\subsubsection{Offloading complexity from end devices}
\label{sec:offloadcomplexity}
%In terms of energy, the end-devices are the most-constrained elements in LPWA networks. Therefore, most technologies simplify the design of end devices by offloading complex tasks to the base stations or to the backend system constituted by network and application servers. Typically, the base stations exploit hardware diversity and are capable of transmitting to and listening from multiple end devices using different channels or orthogonal signals simultaneously. The backend system may be responsible for mechanisms to adapt data rate (e.g., to maximize link capacities),  support mobility and suppress network duplicates. \lorawan is an example of LPWA standards that explicitly makes such choices to conserve energy for the end-devices. 

\memo{ 
Most technologies simplify the design of end devices by offloading complex tasks to the base stations or to the backend system. To keep the transceiver design for end devices simple and low cost, the base stations or backend system have to be more complex. Typically, base stations exploit hardware diversity and are capable of transmitting to and listening from multiple end devices using multiple channels or orthogonal signals simultaneously. This allows end devices to send data using any available channel or orthogonal signal and still reach the base station without need for expensive signaling to initiate communication. By embedding some intelligence in backend system, end devices can further benefit from more reliable and energy efficient last mile communication.  A notable example is \lorawan in which backend system adapts communication parameters (such as data rate/ modulation parameters) to maintain good uplink and downlink connections. Furthermore, backend system is also responsible for providing support for end devices to move across multiple base stations and suppress duplicate receptions if any. The choice of keeping complexity at base stations and backend systems, which are fewer in number, enables low cost and low power design for many end devices. 

Apart from communication, data processing can also be offloaded from end devices but we need to understand a few trade-offs first. Given the diversity of IoT applications, each may have different requirements, in particular the data reporting frequency. There may be some applications which require the end devices to report data frequently (e.g., once every few minutes). At the other extreme we may have applications that require the end devices to report data less frequently~\cite{inteldublinlessfrequent, practical}, perhaps once a day. From an energy consumption perspective, it is a well-known fact that a communication operation consumes more energy than a processing operation. Therefore, a key question that often surfaces is whether to report all the data as it is or carry out some local processing and report the processed result (reduced need for communication). The former approach does not require any significant processing capability at the end device which implies low cost devices can be realized. However in the latter case, depending on the sophistication of processing required, the cost of the end device is likely to go up albeit reducing the energy consumption required to transport the data. The choice between the two is really driven by the underlying business case. Whilst it is always desirable to have low cost end devices especially given the large volumes of devices, it may be beneficial to have some local processing if the communication cost is substantial. Similarly, if the communication cost does not depend on the volume of data (because of flat rate pricing), then it may be beneficial to have simpler end devices. It is also necessary to estimate the costs associated with operating an end device with and without sophisticated processing. In other words, how does the cost stacks up if the end device was to be replaced often due battery depletion caused by frequent communication against deploying a slightly more expensive end device in the first place that communicates less often but does not deplete its battery that often. From a network operator's perspective, it may be desirable to reduce the amount of traffic on their network by local processing on the nodes as this may reduce the likelihood of performance issues. However, this may be undesirable if the operator's business model relies on pricing not based on volume of data.  

The paradigm of processing data closer to the end device, more recently being referred to as edge computing, appears to be gaining popularity as evident from the rise of initiatives such as OpenFog~\cite{openfog}  and Mobile Edge Computing~\cite{etsimec}. Having said this, there is no simple one-size-fits-all binary answer to the problem of whether to transport raw data or to transport the locally processed result. As mentioned earlier, this really boils down to the requirements of the application and the analysis of return on investment (ROI) for those that want to deploy such solutions.  
}

\subsection{Low Cost}
%Providing cheap end-devices  affordable connectivity to is vital to the success of LPWA. 
The commercial success of LPWA networks is tied to connecting a large number of end devices, while keeping the cost of hardware below \$5\memo{~\cite{huaweinbiotwhitepaper, gsmaindustrypaper, sigfox2dollar}} and the connectivity subscription per unit as low as \$1. This affordability enables LPWA technologies to not only address a wide-range of applications, but also compete favorably within the domains where the short-range wireless technologies and the cellular networks are already well-established. LPWA technologies adopt several ways to reduce the capital expenses (CAPEX) and operating expenses (OPEX) for both the end-users and network operators. \memo{The low cost design of end devices is made possible by several techniques some of which are already discussed above in~\ref{sec:lowpower}. Use of star-type (instead of mesh) connectivity, simple MAC protocols, and techniques to offload complexity from end devices enables manufacturers to design simple and therefore low-cost end devices. Some more techniques, mechanisms and approaches are discussed as follows:}

%associated revenue per unit device  have to be low 

\subsubsection{Reduction in hardware complexity}
Compared to the cellular and the short range wireless technologies, LPWA transceivers need to process less complex waveforms. It enables them to reduce transceiver footprint, peak data rates, and memory sizes, minimizing the hardware complexity and thus the cost\memo{~\cite{ericssonwhitepaper}. LPWA chip manufacturers target large number of connected end devices and can also reduce cost with economies of scale. }

\subsubsection{Minimum infrastructure}
Traditional wireless and wired technologies suffer from limited range, requiring dense and therefore an expensive deployment of infrastructure (gateways, power lines, relay nodes etc.). 
%No gateways, wiring, PLC etc. 
However, a single LPWA base station connects tens of thousands of end devices distributed over several kilometers, significantly reducing the costs for network operators. 
%A very wide-area coverage obviates the need for a dense deployment of gateways as in short-range wireless technologies, 
%The infrastrucutre cost 
%The cost of infrastructure longer range than even more than the cellular 
\subsubsection{Using license-free or owned licensed bands}
% In the midst of the looming spectrum crunch,
\usman{avoid licensing fees}
The cost to network operators for licensing new spectrum for LPWA technologies conflicts with low-cost deployment, short time-to-market \memo{and competitiveness of their subscription offers to customers}. Therefore, most LPWA technologies considered deployment in the license-exempt bands including the industrial, scientific and medical (\ism) band or TV-white spaces. %While interference brings its own challenges as discussed in  Section\ref{sec:challenges}, it 
\nbiot, the LPWA standard from 3GPP, \emph{may} share the cellular bands already owned by MNOs to avoid additional licensing cost. However, to get a better performance, a stand-alone licensed band can be acquired as well, a trend proprietary LPWA technologies may eventually follow to avoid performance degradation due to an increase in number of connected devices using shared spectrum. 
%3GPP is making their debut with operating their proposed standard, NB-IoT, in licensed spectrum. 
%
%Performance issues due to increase in size of LPWA network over time may push the industry to evantually acquire licensed spectrum.
\subsection{Scalability}
%regulations:  stringent resirictions  channel access 
\usman{single base station catering ten thousands to hundred of thousands devices}
The support for massive number of devices sending low traffic volumes is one of the key requirements for LPWA technologies. These technologies should work well with increasing number and densities of connected devices. Several techniques are considered to cope up with this scalability problem. 

\subsubsection{Diversity techniques}
To accommodate as many connected devices as possible, efficient exploitation of diversity in channel, time, space, and hardware is vital. Due to low-power and inexpensive nature of the end devices, much of this is achieved by cooperation from more powerful components in LPWA networks such as base stations and backend systems. LPWA technologies employ multi-channel and multi-antenna communication to parallelize transmissions to and from the connected devices. Further, communication is made resilient to interference by using multiple channels and doing redundant transmissions. 
%Use of multi-channel communication 
%Use of multiple modems at base stations 
%Using multiple orthogonal signals for communication from concurrent devices

\subsubsection{Densification}
To cope up with increased density of the end devices in certain areas, LPWA networks, like traditional cellular networks, will resort to dense deployments of base stations. \memo{The problem, however, is to do so without causing too much interference between end devices and densely deployed base stations. Novel densification approaches for LPWA networks need further investigation because existing cellular techniques rely on well-coordinated radio resource management within and between cells, an assumption not true for most LPWA technologies.}

\subsubsection{Adaptive Channel Selection and Data Rate}
\memo {Not only the LPWA systems should scale to number of connected devices, but individual links should be optimized for reliable and energy efficient communication. 
%The overall capacity of the network depends on communication reliability of each individual link. 
Adapting the modulation schemes, selecting better channels to reach distances reliably, or doing adaptive transmission power control require efficient monitoring of link qualities and coordination between end devices and network.

The extent to which adaptive channel selection and modulation is possible depends on the underlying LPWA technology. Different factors such as link asymmetry and maximum allowable radio duty cycle may limit possibility for very robust adaptive mechanisms. In the cases when the base station is unable to give feedback on quality of uplink communication and/or inform the end devices to adapt their communication parameters, the end devices resort to very simplistic mechanism to improve link quality. Such mechanism includes transmitting same packet multiple times often on multiple randomly selected channels in a hope that at least one copy reaches base station successfully. Such mechanisms arguably enhance reliability for this best-effort uplink communication, while keeping the complexity and cost of end devices very low. In the cases when some downlink communication can enable adaptation of uplink parameters, base stations or backend systems can play a vital role in selecting optimal parameters such as channel or optimal data rate to improve reliability and energy efficiency. } 

\memo{In summary, there is a clear trade off between network scalability and simplicity of low cost end devices. Most LPWA technologies let low-power end devices access limited radio resources in mostly uncoordinated and random fashion, limiting the number of devices that can be supported by the networks. Increasing number of recently published studies~\cite{trllorascale, lancasterlorascale, understandinglimitsoflora, capacityscalability} are revealing practical limitations on the scalability of LPWA networks. In Section~\ref{sec:challenges}, we discuss it as an interesting avenue for future research. }

\subsection{Quality of Service}
%TBA
LPWA technologies target diverse set of applications with varying requirements. At one extreme, it caters to delay tolerant smart metering applications, while on other end it should deliver the alarms generated by home security applications in minimum time\usman{different example??}. %While smart metering applications can tolerate long delays, public safety and security applications raise alarms that should be delivered as soon as possible. 
Therefore, network should provide some sort of quality of service (QoS) over the same underlying LPWA technology. 
%same infrastructure shared between a few delay tolerant applications (metering) as well as a short latency bounds (alarms)
For cellular standards where the underlying radio resources may be shared between LPWA and mobile broadband applications, mechanisms should be defined for co-existence of different traffic types. To the best of our knowledge, current LPWA technologies provide no or limited QoS.

%% file: specifications.tex
%\afterpage{%
%    \clearpage% Flush earlier floats (otherwise order might not be correct)
%    \thispagestyle{empty}% empty page style (?)
%    \begin{landscape}% Landscape page
 %       \centering % Center table
\begin{table*}[!t]
\scriptsize
\caption{Technical specifications of various LPWA technologies (?=Not Known)}
\label{tab:specifications}
\vspace{-2mm}
\centering
%\hspace*{-6mm}
%\begin{tabular}{|c|c|c|c|c|c|c|c|c|c|c|}\hline
%\begin{tabulary}{1\textwidth} {|J|J|J|J|J|J|J|J|J|J|J|J| }\hline
\begin{tabular}{ | C{2.0cm} || C{3.5cm} | C{3.5cm} | C{3.5cm} |C{3.5cm} |}\hline
%& \multicolumn{10}{c|}{\textbf{updates sent in a given epoch}}\\
%\cline{2-17}
& \bf{\sigfox}  
& \bf{\lorawan} 
%& \bf{\weightless} 
& \bf{\ingenu} 
& \bf{\telensa} % Telensa 
%& \bf{\qowisio}
%& \dash  
%& IEEE 802.15.4k 
%& IEEE 802.15.4g
%& ETSI LTN 
%& 3GPP \nbiot 
\\\hline\hline
\textbf{Modulation}  %<---------------
& UNB DBPSK\ul, GFSK\dl %\sigfox  
& CSS %\lorawan 
%& UNB(N) %\weightless 
& \rpma-DSSS\ul, CDMA\dl%\ingenu 
& UNB 2-FSK %\telensa
%& UNB %\qowisio
%& GFSK %DASH-7   
%& DSSS, FSK%IEEE 802.15.4k 
%& MR-(FSK,OFDMA,OQPSK)%IEEE 802.15.4g
%& UNB (BPSK\ul\usman{No Listen before talk (LBT) or adaptive frequency agility (AFA)}, GFSK\dl) or OSSS%ETSI LTN 
%& %NB-IoT 
\\\hline
\textbf{Band} %<---------------
& \subghz \ism:EU (868MHz), US(902MHz) %\sigfox  
& \subghz \ism:EU (433MHz 868MHz), US (915MHz), Asia (430MHz) %\lorawan 
%& \subghz \ism or TV whitespaces %\weightless 
& \ism 2.4GHz %\ingenu 
& \subghz bands including \ism:EU (868MHz), US (915MHz), Asia (430MHz) %\telensa
%& %\qowisio
%& \subghz 433MHz, 868MHz, 915MHz %DASH-7   
%& \ism \subghz \& 2.4GHz%IEEE 802.15.4k 
%& \ism \subghz \& 2.4GHz %IEEE 802.15.4g
%& \subghz \ism %ETSI LTN 
%& %NB-IoT 
\\\hline
%\textbf{PHY Multiple Access}  %<--------------- 
\textbf{Data rate} %<---------------
& 100 bps\ul, 600 bps\dl %\sigfox  
& 0.3-37.5 kbps (\lora), 50 kbps (FSK) %\lorawan 
%& %\weightless 
& 78kbps \ul, 19.5 kbps\dl~\cite{comparisontable} \usman{156 kbps per sector with 8 channels but 10 kbps http://www.slideshare.net/MaartenWeyn1/overview-of-low-power-wide-area-networks} %\ingenu
& 62.5 bps\ul, 500 bps\dl  %\telensa 
%& %\qowisio
%& 9.6, 55, 167 kbps %DASH-7   
%& 1.5bps-128kbps %IEEE 802.15.4k 
%& 4.8kbps-800kbps%IEEE 802.15.4g
%& %ETSI LTN 
%& %NB-IoT 
\\\hline

\textbf{\memo{Range}} %<---------------
& \memo{10 km \urban, 50 km \rural } %\sigfox  
& \memo{5 km\urban, 15 km \rural} %\lorawan 
%& %\weightless 
& \memo{15 km \urban} %\ingenu
& \memo{1 km \urban} %\telensa 
%& %\qowisio
%& 9.6, 55, 167 kbps %DASH-7   
%& 1.5bps-128kbps %IEEE 802.15.4k 
%& 4.8kbps-800kbps%IEEE 802.15.4g
%& %ETSI LTN 
%& %NB-IoT 
\\\hline

\textbf{Num. of channels / orthogonal signals}
& 360 channels%\sigfox  
& 10 in EU, 64+8\ul and 8\dl in US plus multiple SFs  %\lorawan 
%& Multiple 200Hz channels (N) %\weightless 
& 40 1MHz channels, up to 1200 signals per channel  %\ingenu 
& multiple channels %\telensa
%& %\qowisio
%& 3 types of channels (number depends on band \&region)%DASH-7   
%& %IEEE 802.15.4k 
%& %IEEE 802.15.4g
%& \nd %ETSI LTN 
%& %NB-IoT 
\\\hline
\textbf{Link symmetry}  %<---------------
& \xmark %\sigfox  
& \cmark %\lorawan 
%& %\weightless 
&  \xmark %\ingenu 
& \xmark %\telensa
%& %\qowisio
%& %DASH-7   
%& %IEEE 802.15.4k 
%& %IEEE 802.15.4g
%& %ETSI LTN 
%& %NB-IoT 
\\\hline
\textbf{Forward error correction} %<---------------
& \xmark %\sigfox  
& \cmark %\lorawan 
%& %\weightless 
& \cmark %\ingenu 
& \cmark %\telensa
%& %\qowisio
%& \cmark %DASH-7   
%& \cmark %IEEE 802.15.4k 
%& \cmark %IEEE 802.15.4g
%& \xmark (error detection only) %ETSI LTN 
%& %NB-IoT 
\\\hline
\textbf{MAC}  %<--------------- 
& unslotted \aloha%\sigfox  
& unslotted \aloha %\lorawan 
%& %\weightless 
& CDMA-like %\ingenu
& ? %\telensa 
%& %\qowisio
%& %DASH-7   
%& %IEEE 802.15.4k 
%& %IEEE 802.15.4g
%& \nd %ETSI LTN 
%& %NB-IoT 
\\\hline
\textbf{Topology}  %<--------------- 
& star%\sigfox  
& star of stars%\lorawan 
%& %\weightless 
& star, tree %\ingenu
& star %\telensa 
%& %\qowisio
%& %DASH-7   
%& %IEEE 802.15.4k 
%& %IEEE 802.15.4g
%& \nd %ETSI LTN 
%& %NB-IoT 
\\\hline

\textbf{Adaptive Data Rate}  %<---------------
& \xmark %\sigfox  
& \cmark %\lorawan 
%& %\weightless 
& \cmark %\ingenu 
& \xmark %\telensa
%& \xmark %\qowisio
%& \xmark %DASH-7   
%& %IEEE 802.15.4k 
%& %IEEE 802.15.4g
%& %ETSI LTN 
%& %NB-IoT 
\\\hline
\textbf{Payload length}  %<---------------
& 12B\ul, 8B\dl %\sigfox  
& up to 250B (depends on SF \& region)%\lorawan 
%& %\weightless 
& 10KB %\ingenu
& ? %\telensa 
%& %\qowisio
%& 256B %DASH-7   
%& 2047B%IEEE 802.15.4k 
%& 2047B%IEEE 802.15.4g
%& 255B %ETSI LTN 
%& %NB-IoT 
\\\hline
\textbf{Handover}  %<---------------
& end devices do not join a single base station %\sigfox  
& end devices do not join a single base station %\lorawan 
%& %\weightless 
& \cmark %\ingenu 
& ? %\telensa
%& %\qowisio
%& \cmark%DASH-7   
%& %IEEE 802.15.4k 
%& %IEEE 802.15.4g
%& \cmark %ETSI LTN 
%& %NB-IoT 
\\\hline
\textbf{Authentication \& encryption}  %<---------------
& encryption not supported %\sigfox  
& AES 128b  %\lorawan 
%& %\weightless 
& 16B hash, AES 256b %\ingenu 
& ? %\telensa
%& %\qowisio
%& AES 128b%DASH-7   
%& AES 128b %IEEE 802.15.4k 
%& AES 128b%IEEE 802.15.4g
%& \cmark %ETSI LTN 
%& %NB-IoT 
\\\hline
\textbf{Over the air updates}  %<---------------
& \xmark %\sigfox  
& \cmark %\lorawan 
%& %\weightless 
& \cmark %\ingenu
& \cmark %\telensa 
%& %\qowisio
%& possible %DASH-7   
%& %IEEE 802.15.4k 
%& %IEEE 802.15.4g
%& \nd %ETSI LTN 
%& %NB-IoT 
\\\hline
%\textbf{Roaming support}  %<---------------
%& %\sigfox  
%& %\lorawan 
%& %\weightless 
%& %\ingenu
%& %\telensa 
%& %\qowisio
%& %DASH-7   
%& %IEEE 802.15.4k 
%& %IEEE 802.15.4g
%& \cmark %ETSI LTN 
%& %NB-IoT 
%\\\hline
\textbf{SLA support}  %<---------------
& \xmark %\sigfox  
& \xmark %\lorawan 
%& %\weightless 
& \xmark %\ingenu
& \xmark %\telensa 
%& \xmark %\qowisio
%& \xmark %DASH-7   
%& %IEEE 802.15.4k 
%& %IEEE 802.15.4g
%& %ETSI LTN 
%& %NB-IoT 
\\\hline
\textbf{Localization}  %<---------------
& \xmark %\sigfox  
& \cmark %\lorawan 
%& \xmark%\weightless 
& \xmark%\ingenu 
& \xmark%\telensa
%& \xmark%\qowisio
%& \xmark%DASH-7   
%& %IEEE 802.15.4k 
%& %IEEE 802.15.4g
%& %ETSI LTN 
%& %NB-IoT 
\\\hline
%\textbf{No. of devices}  %<---------------
%& %\sigfox  
%& %\lorawan 
%%& %\weightless 
%& 16k\usman{from telefonica}%\ingenu 
%& 10k %\telensa
%%& %\qowisio
%%& %DASH-7   
%%& %IEEE 802.15.4k 
%%& %IEEE 802.15.4g
%%& \cmark %ETSI LTN 
%%& %NB-IoT 
%\\\hline
\end{tabular}
%\end{tabulary}
\vspace*{-6mm}
\end{table*}
%       \captionof{table}{Table caption}% Add 'table' caption
%    \end{landscape}
%    \clearpage% Flush page
%}

%% file: standardspecifications.tex
%\afterpage{%
%    \clearpage% Flush earlier floats (otherwise order might not be correct)
%    \thispagestyle{empty}% empty page style (?)
%    \begin{landscape}% Landscape page
 %       \centering % Center table
\begin{table*}[!t]
\scriptsize
\caption{Technical specifications of various LPWA standards}
\label{tab:standardspecifications}
\vspace{-2mm}
\centering
%\hspace*{-6mm}
%\begin{tabular}{|c|c|c|c|c|c|c|c|c|c|c|}\hline
%\begin{tabulary}{1\textwidth} {|J|J|J|J|J|J|J|J|J|J|J|J| }\hline
\begin{tabular}{ | C{1.6cm} || C{2.2cm} | C{2.2cm} | C{2.2cm} |C{2.2cm} | C{2.2cm} | C{2.2cm} |}\hline
%& \multicolumn{10}{c|}{\textbf{updates sent in a given epoch}}\\
%\cline{2-17}
\multirow{2}{*}{\bf{Standard}}
%\bf{SDO/SIG}
& \multicolumn{2}{c|}{\bf{IEEE}} 
& \multicolumn{3}{c|}{\bf{\weightless-SIG}} 
& \bf{\dash Alliance} 
%& eMTC 
%& EC-GSM
%& \nbiot 
\\
%\bf{Standard}
& \bf{802.15.4k} 
& \bf{802.15.4g}
& \bf{\weightless-W} 
& \bf{\weightless-N}
& \bf{\weightless-P}  
& \bf{\dash} 
%& eMTC 
%& EC-GSM
%& \nbiot 
\\\hline\hline
\textbf{Modulation}  %<---------------
& DSSS, FSK %IEEE 802.15.4k 
& MR-(FSK, OFDMA, OQPSK)%IEEE 802.15.4g
& 16-QAM, BPSK, QPSK, DBPSK%\weightless-W 
& UNB DBPSK %\weightless-N
& GMSK, offset-QPSK  %\weightless-P  
& GFSK%\dash  
%& %eMTC 
%& %EC-GSM
%& %\nbiot 
\\\hline
\textbf{Band} %<---------------
& \ism \subghz \& 2.4GHz %IEEE 802.15.4k 
& \ism \subghz \& 2.4GHz %IEEE 802.15.4g
& TV white spaces 470-790MHz%\weightless-W 
& \ism \subghz EU (868MHz), US (915MHz)%\weightless-N
& \subghz \ism or licensed %\weightless-P  
& \subghz 433MHz, 868MHz, 915MHz %\dash  
%& %eMTC 
%& %EC-GSM
%& %\nbiot 
\\\hline
\textbf{Data rate} %<---------------
& 1.5 bps-128 kbps %IEEE 802.15.4k 
& 4.8 kbps-800 kbps%IEEE 802.15.4g
& 1 kbps-10 Mbps%\weightless-W 
& 30 kbps-100 kbps%\weightless-N
& 200 bps-100kbps%\weightless-P  
& 9.6,55.6,166.7 kbps %\dash  
%& %eMTC 
%& %EC-GSM
%& %\nbiot 
\\\hline
\textbf{\memo{Range}} %<---------------
& \memo{5 km \urban} %IEEE 802.15.4k 
& \memo{up to several kms} %IEEE 802.15.4g
& \memo{5 km \urban} %\weightless-W 
& \memo{3 km \urban} %\weightless-N
& \memo{2 km \urban} %\weightless-P  
& \memo{0-5 km \urban}  %\dash   0-5 km ???
%& %eMTC 
%& %EC-GSM
%& %\nbiot 
\\\hline

%\textbf{PHY Multiple Access}  %<--------------- 
\textbf{Num. of channels / orthogonal signals}  %<--------------- 
&\multicolumn{2}{C{4.4cm}|}{multiple channels. \newline Number depends on channel \& modulation}
%& %IEEE 802.15.4k 
%& multiple channels. Number depends on channel \& modulation%IEEE 802.15.4g
& 16 or 24 channels\ul %\weightless-W 
& multiple 200 Hz channels %\weightless-N
& multiple 12.5 kHz channels%\weightless-P  
& 3 different channel types (number depends on type \& region) %\dash  
%& %eMTC 
%& %EC-GSM
%& %\nbiot 
\\\hline
%\textbf{Link symmetry}  %<---------------
%& \cmark %IEEE 802.15.4k 
%& \cmark %IEEE 802.15.4g
%& %\weightless-W 
%& \xmark (\ul only)%\weightless-N
%& \cmark %\weightless-P  
%& \cmark %\dash  
%%& %eMTC 
%%& %EC-GSM
%%& %\nbiot 
%\\\hline

\textbf{Forward error correction} %<---------------
& \cmark%IEEE 802.15.4k 
& \cmark%IEEE 802.15.4g
& \cmark %\weightless-W 
& \xmark %\weightless-N
& \cmark%\weightless-P  
& \cmark%\dash  
%& %eMTC 
%& %EC-GSM
%& %\nbiot 
\\\hline
\textbf{MAC}  %<--------------- 
& CSMA/CA, CSMA/CA or \aloha with PCA%IEEE 802.15.4k 
& CSMA/CA %IEEE 802.15.4g
& TDMA/FDMA%\weightless-W 
& slotted \aloha%\weightless-N
& TDMA/FDMA%\weightless-P  
& CSMA/CA %\dash  
%& %eMTC 
%& %EC-GSM
%& %\nbiot 
\\\hline
\textbf{Topology}  %<--------------- 
& star %IEEE 802.15.4k 
& star, mesh, peer-to-peer (depends on upper layers) %IEEE 802.15.4g
& star%\weightless-W 
& star %\weightless-N
& star %\weightless-P  
& tree, star %\dash  
%& %eMTC 
%& %EC-GSM
%& %\nbiot 
\\\hline
%\textbf{Adaptive Data Rate}  %<---------------
%& %IEEE 802.15.4k 
%& %IEEE 802.15.4g
%& %\weightless-W 
%& %\weightless-N
%& %\weightless-P  
%& %\dash  
%%& %eMTC 
%%& %EC-GSM
%%& %\nbiot 
%\\\hline
\textbf{Payload length}  %<---------------
& 2047B %IEEE 802.15.4k 
& 2047B %IEEE 802.15.4g
& $>$10B%\weightless-W 
& 20B %\weightless-N
& $>$10B%\weightless-P  
& 256B %\dash  
%& %eMTC 
%& %EC-GSM
%& %\nbiot 
\\\hline
%\textbf{Over-the-air updates}  %<---------------
%& %IEEE 802.15.4k 
%& %IEEE 802.15.4g
%& %\weightless-W 
%& %\weightless-N
%& \cmark %\weightless-P  
%& possible%\dash  
%%& %eMTC 
%%& %EC-GSM
%%& %\nbiot 
%\\\hline
\textbf{Authentication \& encryption}  %<---------------
& AES 128b%IEEE 802.15.4k 
& AES 128b%IEEE 802.15.4g
& AES 128b %\weightless-W 
& AES 128b %\weightless-N
& AES 128/256b%\weightless-P  
& AES 128b%\dash  
%& %eMTC 
%& %EC-GSM
%& %\nbiot 
\\\hline
%\textbf{Handover}  %<---------------
%& %IEEE 802.15.4k 
%& %IEEE 802.15.4g
%& %\weightless-W 
%& %\weightless-N
%& %\weightless-P  
%& \cmark %\dash  
%%& %eMTC 
%%& %EC-GSM
%%& %\nbiot 
%\\\hline
%%\textbf{Roaming support}  %<---------------
%%& %IEEE 802.15.4k 
%%& %IEEE 802.15.4g
%%& %\weightless-W 
%%& %\weightless-N
%%& %\weightless-P  
%%& %\dash  
%%%& %eMTC 
%%%& %EC-GSM
%%%& %\nbiot 
%%\\\hline
%\textbf{SLA support}  %<---------------
%& %IEEE 802.15.4k 
%& %IEEE 802.15.4g
%& %\weightless-W 
%& %\weightless-N
%& %\weightless-P  
%& %\dash  
%%& %eMTC 
%%& %EC-GSM
%%& %\nbiot 
%\\\hline
%\textbf{Localization}  %<---------------
%& %IEEE 802.15.4k 
%& %IEEE 802.15.4g
%& %\weightless-W 
%& %\weightless-N
%& %\weightless-P  
%& %\dash  
%%& %eMTC 
%%& %EC-GSM
%%& %\nbiot 
%\\\hline
%\textbf{No. of devices}  %<---------------
%& %IEEE 802.15.4k 
%& %IEEE 802.15.4g
%& %\weightless-W 
%& %\weightless-N
%& %\weightless-P  
%& %\dash  
%%& %eMTC 
%%& %EC-GSM
%%& %\nbiot 
%\\\hline
\end{tabular}
%\end{tabulary}
\vspace*{-6mm}
\end{table*}
%       \captionof{table}{Table caption}% Add 'table' caption
%    \end{landscape}
%    \clearpage% Flush page
%}

%% file: business_table.tex
%\afterpage{%
%    \clearpage% Flush earlier floats (otherwise order might not be correct)
%    \thispagestyle{empty}% empty page style (?)
%    \begin{landscape}% Landscape page
 %       \centering % Center table
\begin{table*}[!t]
\scriptsize
\caption{Business considerations for various LPWA technologies (?=Not Known)}
\label{tab:business}
\vspace{-2mm}
\centering
%\hspace*{-6mm}
%\begin{tabular}{|c|c|c|c|c|c|c|c|c|c|c|} \hline
\begin{tabular}{ | C{2cm} | C{2.7cm} | C{2.7cm} | C{2.7cm} |C{2.7cm} | C{2.7cm} | C{2.8cm} |}\hline
%\begin{tabulary}{1\textwidth} {|J||J|J|J|J|J| }\hline
%& \multicolumn{10}{c|}{\textbf{updates sent in a given epoch}}\\
%\cline{2-17}
& \bf{\sigfox} \cite{sigfox}
& \bf{\lorawan}  \cite{loraalliance}
& \bf{\weightless-N} \cite{weightless}
& \bf{\ingenu}\cite{lightreadingingenu}
& \bf{3GPP Cellular IoT}
\\\hline\hline

\textbf{Deployment model}  %<---------------
& Nationwide (multiple countries) %\sigfox
& Private or nationwide networks %\lorawan
& Private networks %\weightless
& Private or nationwide networks %\ingenu
& Nationwide networks %cellular
\\\hline

\textbf{Ease of roaming}  %<---------------
& Seamless roaming across \sigfox networks in different countries at no extra charges %\sigfox
& Roaming agreements required %\lorawan
& Not applicable %\weightless
& ?  %\ingenu
& Operator alliances for cross-border roaming %cellular
\\\hline

\textbf{SLA support}  %<---------------
& \xmark %\sigfox
& \xmark %\lorawan
&  \xmark %\weightless
&  \xmark %\ingenu
&  \cmark %cellular
\\\hline

\textbf{Device availability}  %<---------------
& \cmark %\sigfox
& \cmark %\lorawan
& Focus is on gateway %\weightless
& \cmark %\ingenu
& \xmark (Still in standardization phase, devices will emerge later)%cellular
\\\hline

\textbf{Over-the-air updates for devices}  %<---------------
& \xmark %\sigfox
& possible %\lorawan
& \xmark %\weightless
&  possible %\ingenu
& likely be made available %cellular
\\\hline

\textbf{Supplier ecosystem}  %<---------------
& Transceivers and modules from many vendors %\sigfox
& Limited choice of vendors for transceivers, several module vendors %\lorawan
& Limited choice of vendors %\weightless
& Transceivers and modules from many vendors %\ingenu
& Availability likely from all the usual vendors once standard is ratified %cellular
\\\hline

\textbf{Licensing}  %<---------------
& Technology freely available for chip/device vendors. Network operators pay royalty to \sigfox (revenue sharing basis) %\sigfox
& Technology licensed by device vendors. No royalty to be paid by network operators %\lorawan
& Technology freely available for chip/device vendors. No royalty thereafter. %\weightless
& Upfront fee + per application \& per device fee / year (No revenue sharing) %\ingenu
& Standardized technology. Usual cellular model likely to prevail %cellular
\\\hline

\textbf{Deployment status}  %<---------------
& Network deployed \& running in several countries.  Several operators have invested in \sigfox %\sigfox
& Early trials \& deployments by some operators. Several operators are members of \loraalliance %\lorawan
& Some trials but no major deployments %\weightless
& Several private deployments in over 5 continents %\ingenu
& Early days with some in-house trials with pre-standardized technology by handful of operators  %cellular
\\\hline

\textbf{Longevity offered by the solution}  %<---------------
& Deployments in several countries. Not much insight into transition plan should SNOs find it infeasible to run the network. Transitioning will entail replacement of endpoint/communications module in the endpoints. %\sigfox
& Some deployments by cellular operators in a few countries. No insight into transition plan should LoRa network be decommissioned. Transitioning will entail replacement of endpoint/communications module in the endpoints. %In the event of de-commisioning \lorawan customers could be migrated to one of the cellular based IoT alternatives %\lorawan
& No deployments so far so longevity is questionable   %\weightless
& Deployments in several countries. Not much insight into transition plan should MNOs find it infeasible to run the network. Transitioning will entail replacement of endpoint/communications module in the endpoints. %\ingenu
& Promising as this being a solution designed exclusively for IoT, is less likely to be de-commissioned.  %cellular
\\\hline

%\textbf{Ease of integration with IoT platforms} %<---------------
%& Partnerships with most of the big IoT platform providers %\sigfox
%& Not as extensive support for IoT platforms as \sigfox \usman{is it true?} %\lorawan
%& No information available   %\weightless
%& Expect to be supported by most IoT platform providers once products are available.  %cellular
%\\\hline

\end{tabular}
%\end{tabulary}
\vspace*{-6mm}
\end{table*}
%       \captionof{table}{Table caption}% Add 'table' caption
%    \end{landscape}
%    \clearpage% Flush page
%}

%% file: business.tex
\section{Business Considerations}
\label{sec:business}

%\usman{Parag, the next paragraph contains your text. Please see some commented out pointers in the latex as well}
%\usman{many proprietary technologies, not much research community contribution, required for solving big challenges, interoperability link 2 independent, no comparison of different technology in real settings}

%STUNNING NUMBERS: 26 billion connected devices in 2020 among which 17 billion are consumer electronics and M2M devices
%1.9 trillion dollar value add of Iot across different sectors in 2020.
%MANY APPLICATIONS ( Agriculture , Transportation, Energy, Health care): smart city, sustainable agriculture, industrial monitoring, transportation, health care, logistics tracking and fleet management, smart meter, buildings, smart meters water gas

With the dawn of the M2M communications paradigm, 2G seemed to be a reasonable fit for catering to the requirements of these applications. Given the spectrum scarcity worldwide and the high capital expenditure incurred in acquiring new spectrum, operators appear to be in a dilemma whether to continue using 2G systems for serving M2M customers or re-farm the spectrum making way for new technologies such as LTE and its variants. Announcements from a handful of operators to transition to the latter created a hole in the market. Since then, several new LPWA technologies such as those mentioned in Table \ref{tab:business} have been aggressively trying to fill this gap with the hope of staking their claim to the pole position. Only those technologies have been included in the table for which substantial information is available in the public domain, those that have a wide variety of products already available in the market and those that have had large scale deployments. %Ingenu was excluded as there was hardly any information available (as of this writing) to compare it against the rest. 
The much anticipated NB-IoT standard from the cellular world has been included to provide a perspective as to how the different forerunners in the market stack up against a potential cellular offering in the making.

It is worth emphasizing that there is no one size fits all solution with each of these approaches having their pros and cons as highlighted in the table. The market is still up for grabs and players have several strategic options to consider depending on their circumstances, e.g., those needing to deploy an IoT solution immediately will have to hedge their bets on \lora, \sigfox, \ingenu, \weightless-N etc. whereas others can afford to wait until the 3GPP finalizes standards such as \nbiot which is still work-in-progress. In the meanwhile, the cellular operators themselves seem to have hedged their bets on \lora and \sigfox with several operators making big investments in one or the other. In any case, it looks like a win-win situation for the operators irrespective of how the situation plays out since these technologies could play a complementary role to the potential NB-IoT standard that is currently being baked. Also, the fact that operators have invested in these technologies reduces the uncertainty from a longevity perspective\footnote{Recall the nightmares that the announcements regarding sunset of the 2G systems might have given to the M2M customers} for the adopters of these solutions. 

It is envisaged that \lora, \sigfox, and \ingenu will continue to challenge the hegemony of the cellular players and all four are likely to share the pie in the long run. It is expected that there would be a varying degree of adoption across multiple market segments and pricing models \cite{rethinkiot} are likely to have a significant impact on the success of different technologies. 

\memo{In a nutshell, as of this writing fierce battles continue to be fought to capture the LPWA market share and competitors are leaving no stone unturned to attack each other's propositions (see \cite{loradefends, vodafonecrush, ingenurevs})}.

%\usman{Parag, I have put below some random thoughts in an unstructured text that I believe you can probably use to complete this section.}

%LPWA good for monitoring static or mobile entities spread over large geographical areas, requiring low data rate.
%However, in closed spaces such as building, adoption is expected to be low because of their competition with established technologies like ZigBee, WLAN etc.. Low data rate of LPWA is also a negative point there.

%For industrial control and automation, actuation the LPWA technology is not optimized. nor exiting technologies target it as their business. LPWA is likely not to meet Ultra-low latency and high reliability requirements, typical of industrial control and automation.

% ALREADY ADOPTED IN: Sample applications: livestock, monitoring pregnancy for livestock, mailboxes, waste bins, agricultural farms, city lighting and natural disasters. The tracking applications for freight, elderly, kids, patients, pets and bikes were also a subject of many talks.

%Business Angle/ Pricing Models will affect the success of different technologies. End consumers only vs. Enterprise specific networks e.g., the two options provided by \lora)
%
%Capabilities of different LPWA are different and so are their expected success rates for certain applications: range \lora $>$ \sigfox in EU and USA but its capacity (number of devices per base station ) is less.
%Weightless-P blends competitive features

%A very good reference:  \url{http://www.weightless.org/about/mobile-experts-group-lpwan-report/NTZiMy9tZXhwLWxwd2EtMTYgZXhlYyBzdW1tYXJ5LnBkZg==}
%
%\url{http://www.gsma.com/connectedliving/wp-content/uploads/2016/03/Mobile-IoT-Low-Power-Wide-Area-Connectivity-GSMA-Industry-Paper.pdf}
%
%\url{http://www.edn.com/design/systems-design/4440343/2/Low-power-wide-area-networking-alternatives-for-the-IoT}
%
%\url{https://docs.google.com/document/d/1n7cXN4_VuI8imy8MG3-fHjl9FNiNvYfdB4txN4hDQ-w/edit#}
%
%
%\url{http://www.eejournal.com/archives/articles/20150907-lpwa/} 
%
%\url{http://rethink-iot.com/2015/03/20/on-lpwans-why-sigfox-and-lora-are-rather-different-and-the-importance-of-the-business-model/}